\documentclass[twocolumn]{autart}    

\usepackage{cite}
\usepackage{amsmath,amssymb,amsfonts}
\usepackage{algorithmic}
\usepackage{graphicx}
\usepackage{textcomp}
\PassOptionsToPackage{version=3}{mhchem}
\usepackage{mhchem}
\usepackage{chemarr}
\usepackage{verbatim} 
\usepackage{epstopdf, color} 
\usepackage{arydshln}
\usepackage{marginnote}
\usepackage{stmaryrd}

\newtheorem{lemma}{Lemma}
\newtheorem{proposition}{Proposition}
\newtheorem{remark}{Remark}
\newtheorem{assumption}{Assumption}
\newtheorem{example}{Example}

\usepackage{flushend}
\usepackage{balance}

\newif\ifarxiv
\newcommand{\commentarxiv}[1]{\ifarxiv{ #1}\fi}
\newcommand{\myclearpage}{\ifarxiv \else  \fi}

\arxivtrue 

\begin{document}

\begin{frontmatter}
	\title{On nucleic acid feedback control systems} 
	
	\thanks[footnoteinfo]{Corresponding author}

 \author[warwickaddress]{Nuno M. G. Paulino\thanksref{footnoteinfo}}
  \ead{N.Paulino@warwick.ac.uk}
  \author[conventryaddress]{Mathias Foo}
 \ead{Mathias.Foo@coventry.ac.uk}
  \author[postechaddress]{Jongmin Kim}
 \ead{jongmin.kim@postech.ac.kr}
  \author[warwickaddress]{Declan G. Bates}
 \ead{D.Bates@warwick.ac.uk}

 \address[warwickaddress]{Warwick Integrative Synthetic Biology Centre, School of Engineering, University of Warwick, Coventry CV4 7AL, UK}
 \address[conventryaddress]{School of Mechanical, Aerospace and Automotive Engineering, Coventry University, Coventry CV1 5FB, UK}
 \address[postechaddress]{Department of Integrative Biosciences and Biotechnology, Pohang University of Science and Technology (POSTECH), Pohang, Gyeongbuk, 37673, South Korea}

\begin{abstract}

Recent work has shown how chemical reaction network theory may be used to design dynamical systems that can be implemented biologically in nucleic acid-based chemistry. While this has allowed the construction of advanced open-loop circuitry based on cascaded DNA strand displacement (DSD) reactions, little progress has so far been made in developing the requisite theoretical machinery to inform the systematic design of feedback controllers in this context. 
Here, we develop a number of foundational theoretical results on the equilibria, stability, and dynamics of nucleic acid controllers. In particular, we show that the implementation of feedback controllers using DSD reactions introduces additional nonlinear dynamics, even in the case of purely linear designs, e.g. PI controllers.
By decomposing the effects of these non-observable nonlinear dynamics, we show that, in general, the stability of the linear system design does not necessarily imply the stability of the underlying chemical network, which can be lost under experimental variability when feedback interconnections are introduced.
We provide an in-depth theoretical analysis of an example illustrating this phenomenon, whereby the linear design does not capture the instability of the full nonlinear system implemented as a DSD reaction network, and we further confirm these results using VisualDSD, a bespoke software tool for simulating nucleic acid-based circuits. Our analysis highlight the many interesting and unique characteristics of this important new class of feedback control systems.

\end{abstract}


\end{frontmatter}

\myclearpage

\section{Introduction}
\label{sec:introduction}
Recent advances in synthetic biology have seen the incorporation of many control engineering design principles into the construction of biomolecular circuits~\cite{Qian2018,Blanchini,Hancock2019,Siami2019}.
%
One of the current urgent needs of this aspect of synthetic biology is the development of bespoke feedback control theory that can be used to systematically design synthetic controllers for biomolecular processes. 
A promising direction for this work is to exploit chemical reaction network (CRN) theory, since CRN's act as a ``bridge" between mathematical designs based on ordinary differential equations (ODEs) and biological implementations in nucleic acid-based chemistry using DNA strand displacement (DSD) reactions~\cite{Soloveichik2010,Chen2013}.  The capability of such circuits to operate \emph{in vivo} and interface with endogenous cellular machinery has been demonstrated in mammalian cells, with some notable examples including engineered oligonucleotide AND gates responding to microRNA inputs~\cite{Hemphill2013}, multi-input logic based on DNA circuitry interacting with native mRNA~\cite{Groves2015}, and reliable strand displacement probes triggered by mRNA being transcribed into cells~\cite{Chatterjee2018}. This makes circuits based on nucleic acids strong potential candidates for implementing many computing and control applications in synthetic biology.

The CRN to DNA design framework~\cite{Soloveichik2010} assigns a formal species in the CRN to sets of DNA species, allowing the construction of circuits supported by a high level of automation using available syntax and software tools \cite{Phillips2009,lakin2011visual,Yordanova}. 
%
In the context of feedback control, however, a key challenge with employing CRNs is their inability to directly represent negative signals (since concentrations of chemical species are always positive). For example, CRNs generally can only compute a positive difference between two positive inputs, i.e.~``one-sided" subtraction \cite{Buisman2008}.
The use of the so-called \emph{dual-rail representation} with nucleic-acids~\cite{Seelig2006} circumvents this problem by representing each signal as the difference of concentrations of two different species. 
Although it doubles the number of required reactions, the dual rail representation
enables the computation of rational linear functions as the steady state of a CRN~\cite{Chen2014a}, including two-sided subtraction. 
It provides an Internally Positive Representation (IPR), where a positive state-space system, together with input, state and output transformations, can realize arbitrary input/output dynamics~\cite{Cacace2012}.
Unimolecular reactions of catalysis and degradation, and bimolecular reactions of annihilation, can then be used to construct CRNs to approximately represent transfer functions~\cite{Chiu2015},  linear feedback systems~\cite{Oishi2011,Yordanova, Paulino2019a}, and nonlinear controllers~\cite{Sawlekar2016}.

In all these systems, annihilation reactions operating with very fast timescales are essential, in order to ensure that species concentrations remain within the bounds of experimental feasibility. However, these reactions result in a nonlinear IPR.
As noted in~\cite{Paulino2019}, these annihilation reactions introduce additional internal nonlinear dynamics that are not observable in the represented input/output linear dynamics, but become important in the presence of inevitable experimental variability in the biomolecular implementations.
Here, we formally characterise the effects of the nonlinear dynamics introduced through these annihilation reactions on the equilibria and stability of closed-loop nucleic acid systems. These results provide many useful insights that can guide the design and construction of these circuits \emph{in vitro} and \emph{in vivo}, and also highlight some of the associated technical challenges and limitations.

\subsection{Notation and Preliminaries}

%
%
We represent the elements of vectors and matrices $\mathbf{x=Mv}$ with $x_{j}=\left[\mathbf{Mv}\right]_{j}=\sum_{i}m_{ji}v_{i}$. 
$\mathbf{1}$ is a vector with elements $1$, and $\mathbf{I}$ is the identity matrix.
The element-wise product is represented with  $\mathbf{x}=\mathbf{v}\circ\mathbf{u}\Rightarrow x_{j}=v_{j}u_{j}$. For a vector $\mathbf{v}\geq 0$, $\|\mathbf{v}\|_1=\mathbf{1}^T\mathbf{v}$ and $\|\mathbf{v}\|_2^2=\mathbf{1}^T\left(\mathbf{v}\circ\mathbf{v}\right)$.
In the system dynamics, for brevity,  time dependency is implicit, i.e.~$x_j\equiv x_j(t)$, $x_j^*\equiv x_j(\infty)$ represents steady state conditions, and $X_j(s)$ is the Laplace transform of $x_j$.
$\rho\left\{ \mathbf{M}\right\}$ denotes the set of the eigenvalues $\lambda_i$ of matrix $ \mathbf{M}$.
We represent the set of Hurwitz matrices with $\mathcal{H}$. Given the spectral abscissa $\alpha\left\{ \mathbf{M}\right\}=\max_i \Re\left\{\lambda_i\right\}$, if $\mathbf{M}\in\mathcal{H}$, then $\alpha\left\{ \mathbf{M}\right\}<0$.
Given the set of lower triangular matrices $\mathcal{L}$, then for $\mathbf{M}\in\mathcal{L}$ we have that $m_{ji}=0,i > j$, and $\lambda_i\left\{\mathbf{M}\right\}=m_{ii}$. 
Given the set $\mathcal{I}$ of irreducible matrices~\cite{Farina2000}, if $\mathbf{M}\in\mathcal{I}$, then there is no permutation such that $\mathbf{M}\in\mathcal{L}$. Also, if $\mathbf{M}\in \mathcal{L}$, then $\mathbf{M}\notin \mathcal{I}$.
%
%

%
$\mathbb{R}_{0}^{+}$ is the positive orthant, where all the coordinates
of a vector $v_{j}\geq0$. 
$\mathbf{M}\geq0$ means all elements $m_{ji}\geq0$, and $\mathbf{M}\in\mathbb{R}_{0}^{+}$. 
%
%
The operator $\mathbf{D}\left\{ \mathbf{v}\right\} $ is defined as a
square matrix where $d_{jj}=v_j$ and $d_{ji}=0,\, j\neq i$. 
%
%
If $\mathbf{m}$ is the diagonal of $\mathbf{M}$, the matrix of off-diagonal elements $\mathbf{M}^{\boxbslash}$ is defined as $\mathbf{M}^{\boxbslash}=\mathbf{M} - \mathbf{D}\left\{ \mathbf{m}\right\}$.
Defining $\mathcal{M}$ as the group of Metzler matrices, if $\mathbf{M}\in\mathcal{M}$, then 
$\mathbf{M}^{\boxbslash}\geq 0$ and if $\mathbf{M}\in\mathcal{M,H}$ then $\mathbf{m}<0$.

\begin{figure}[!t]
	\centerline{\includegraphics[width=1\columnwidth]{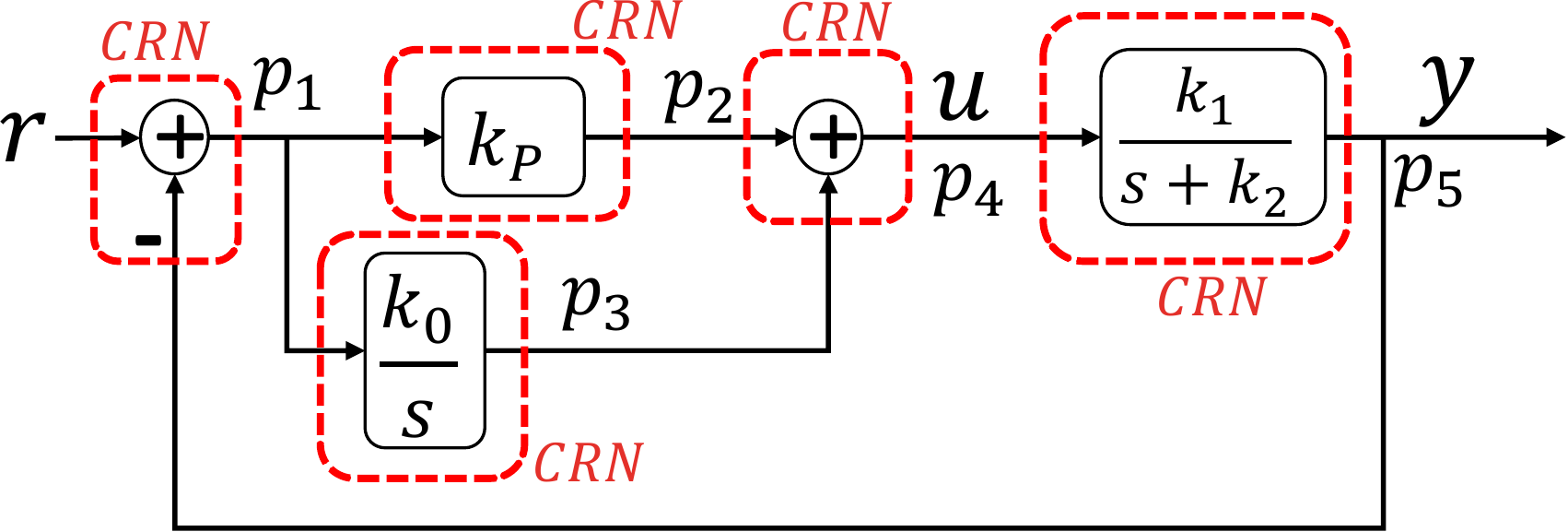}}
	\caption{Linear negative feedback interconnection, where each signal $p_{j}$ results from representing a linear operator with chemical reactions.}
	\label{fig: Feedback-interconnection-linear}
\end{figure}
A CRN is composed of a set of reactions between chemical species $X_{j}$, at a rate $\gamma$. 
The dynamics of a chemical reaction can be approximated by ODEs using mass action kinetics~\cite{PeterToth;JanosErdi1989}, i.e.,
\begin{eqnarray}
&&a_{1}X_{1}+a_{2}X_{2}\xrightarrow{\gamma } b X_{3} \Rightarrow\dot{x}_{3}=b \gamma  x_{1}^{a_{1}}x_{2}^{a_{2}}\label{eq:CRN}
\end{eqnarray}
The stoichiometric coefficients $a_1$, $a_2$ and $b$ indicate, respectively, the relative number of molecules consumed and produced during the reaction.
Note that expressing the dynamics of \eqref{eq:CRN} in their natural coordinates, the concentrations, results in a non-negative state vector $\mathbf{x}\in\mathbb{R}_{0}^{+}$. 
The use of chemical concentrations as state variables is therefore not suitable for circuits involving negative signals, such as feedback control loops.
To circumvent the above problem, it is now standard practice to represent both positive and negative signals with a dual-rail representation, where each signal is the difference between two positive quantities. Consider, for example, the linear feedback control system depicted in Fig.~\ref{fig: Feedback-interconnection-linear}. In the dual-rail representation, all signals in this system are split into contributions from two molecular concentrations $p_{j}=x_{j}^{+}-x_{j}^{-}$, where $x_j^+\geq 0$, $x_j^-\geq 0$, and $p_j\in\mathbb{R}$.
%
Linear mathematical operators and transfer functions can be represented with elementary CRNs~\cite{Oishi2011,Chiu2015} using unimolecular reactions of catalysis and degradation, and bimolecular reactions of annihilation, which are respectively given by
%
\begin{eqnarray}
&&X_i\xrightarrow{\gamma }X_i+X_{j},\quad X_{j}\xrightarrow{\gamma}\emptyset,\quad X_{i}+X_{j}\xrightarrow{\eta}\emptyset\label{crn: elementary CRN}
\end{eqnarray} 
Employing the dual rail representation entails duplicating all the catalysis and degradation reactions. In the following we compact the notation so that $X^\pm$ represents simultaneously both species $X^+$ and $X^-$, and $x^\pm$ their respective two concentrations $x^+$ and $x^-$. $Y^{\pm}\xrightarrow{\gamma^\pm}Y^{\pm}+X^{\mp}$ is an abbreviation for the two parallel reactions  $Y^{+}\xrightarrow{\gamma^+}Y^{+}+X^{-}$ and $Y^{-}\xrightarrow{\gamma^-}Y^{-}+X^{+}$. 

%
%
%

\section{Representation of linear feedback control systems with chemical reaction networks}\label{sec: dual rail CRN}

Each reaction in~\eqref{crn: elementary CRN} has equivalent representation with DSD reactions, and the sets of these reactions can be systematically converted to implementable reactions based on nucleic acids (for example with~\cite{Soloveichik2010}).
Since the dual representation admits infinite combinations of $x_j^+$ and $x_j^-$ for the same difference $p_j=x_j^+ -x_j^-$, in practice, the annihilation reaction in~\eqref{crn: elementary CRN} is used to ensure one of the concentrations is kept close to zero, and  $p_j\approx x^+_j$ or $p_j\approx - x_j^-$.
A rate $\eta$ for the annihilation reactions that is on a much faster timescale than the dynamics of the system is used to keep the concentrations of all molecular species low (i.e. experimentally feasible) even in the presence of transients.
\begin{assumption}\label{assum: rates are equal}
	The nominal parameterisation and nominal implementation assume perfectly designed reaction rates in the absence of variability, and a symmetrical parametrisation where the reaction rates are the same in each dual reaction with $\gamma_j^+=\gamma_j^-=\gamma_j$.
	\end{assumption}
\begin{defn}\label{defn: io dynamics}
	The Input-Output (I/O) dynamics are the response $Y(s)=G(s)R(s)$ from $r=\left(r^{+}-r^{-}\right)$ to $y=\left(y^{+}-y^{-}\right)$, with $r,y\in\mathbb{R}$ and $r^\pm,y^\pm\in\mathbb{R}^+_0$.  
\end{defn}
For example, in Fig.~\ref{fig: Feedback-interconnection-linear} we represent the plant with the following set of chemical reactions
\begin{eqnarray}
X_{4}^{\pm} \xrightarrow{k_{1}^\pm}X_{4}^{\pm}+X_{5}^{\pm}\label{eq: CRN plant 1},\  
X_{5}^{\pm}\xrightarrow{k_{2}^\pm}\emptyset,\ 
X_{5}^{+}+X_{5}^{-}\xrightarrow{\eta}\emptyset\label{crn: plant}
\end{eqnarray}
Computing the I/O dynamics from Definition~\ref{defn: io dynamics} under Assumption~\ref{assum: rates are equal} ($k_i^+=k_i^-=k_i$), the nonlinear terms cancel out and the I/O dynamics of~\eqref{eq: CRN plant 1} represent a first-order linear system $Y(s)=k_{1}\left(s+k_{2}\right)^{-1}U(s)$, with $y=x_5^+-x_5^-$ and $u=x_4^+-x_4^-$. 
%
The use of bimolecular reactions results in an IPR of a linear system based on nonlinear internal positive dynamics, in contrast to IPRs based on linear positive dynamics~\cite{Cacace2012}.

The linear feedback system in Fig.~\ref{fig: Feedback-interconnection-linear} is represented by combining~\eqref{crn: plant} with the CRNs of the linear operations of integration $\dot{p}_{3}=k_{0}p_{1}$, gain $p_{2}=k_Pp_{1}$, and summation $p_{4}=p_{2}+p_{3}$.
Despite the positivity of the concentrations, the dual rail representation allows us to represent the error $p_{1}=r-p_{5}$ with a CRN (see e.g.~\cite{Oishi2011,Yordanova,Paulino2019a}).

%
%
%
\begin{example}\label{exmp: ODEs CRN}
	 The mass action kinetics of the linear feedback structure in Fig~\ref{fig: Feedback-interconnection-linear} results in the following dynamics
	\begin{subequations}
	\begin{eqnarray}
	\dot{x}_{1}^{\pm}&=&-\gamma^\pm_{3} x_{1}^{\pm}+\gamma^\mp_{2}x_{5}^{\mp}+\gamma^\pm_{1}r^{\pm}-\eta x_{1}^{+}x_{1}^{-} \label{eq: crn dyn nonlin first}\\
	\dot{x}_{2}^{\pm}&=&\gamma^\pm_{4}x_{1}^{\pm}-\gamma^\pm_{5}x_{2}^{\pm}-\eta x_{2}^{+}x_{2}^{-} \label{crn: gain}\\
	\dot{x}_{3}^{\pm}	&=&k^\pm_{0}x_{1}^{\pm}-\eta x_{3}^{+}x_{3}^{-}\\
	\dot{x}_{4}^{\pm}	&=&\gamma^\pm_{6}x_{2}^{\pm}+\gamma^\pm_{7}x_{3}^{\pm}-\gamma^\pm_{8}x_{4}^{\pm}-\eta x_{4}^{+}x_{4}^{-}\label{crn: sum}\\
	\dot{x}_{5}^{\pm} & =&k^\pm_{1}x_{4}^{\pm}-k^\pm_{2}x_{5}^{\pm}-\eta x_{5}^{+}x_{5}^{-}\label{eq: crn dyn nonlin last}
	\end{eqnarray}	
	\end{subequations}
Under Assumption~\ref{assum: rates are equal}, the ODEs for $\dot{p}_j=\dot{x}^+_j-\dot{x}^-_j$ and $r=r^+-r^-$ result in the I/O system 
\begin{eqnarray}
&&\dot{\mathbf{p}}=\mathbf{A}_{p}\mathbf{p}+\mathbf{B}_{p}r,\quad
y=p_5 \label{eq: I/O system state space}\\
&&
\mathbf{A}_p=\left[\begin{array}{ccccc}
-\gamma_{3} & 0 & 0 & 0 & -\gamma_{2}\\
\gamma_{4} & -\gamma_{5} & 0 & 0 & 0\\
k_{0} & 0 & 0 & 0 & 0\\
0 & \gamma_{6} & \gamma_{7} & -\gamma_{8} & 0\\
0 & 0 & 0 & k_{1} & -k_{2}
\end{array}\right],\,
\mathbf{B}_p=\left[\begin{array}{c}
\gamma_{1}\\
0\\
0\\
0\\
0
\end{array}\right]
\end{eqnarray}
%
%
\end{example}
\begin{remark}
	The I/O dynamics are an approximation to the original linear system in Fig.~\ref{fig: Feedback-interconnection-linear}, since the representations of subtraction, gain, and sum are exact only at steady state. The impact of the respective transient dynamics~(\ref{eq: crn dyn nonlin first}), (\ref{crn: gain}), and (\ref{crn: sum}) can be mitigated by increasing $\gamma_j$.
\end{remark}

%
%
%

\section{Dynamics of the chemical reaction network }
We now define the class of systems analysed in this work, where we retain the natural non-negative coordinates, so that the states are
the species concentrations $x_{j}^{\pm}$, and the input vector contains both positive and negative components for the reference $\mathbf{r}=\left[r^{+},r^{-}\right]^{T}$,
$r^{\pm}\in\mathbb{R}_{0}^{+}$. 
\begin{assumption}\label{R11 stable}
	Assume the dynamics we wish to represent result in stable I/O dynamics, and therefore $\mathbf{A}_{p}\in\mathcal{H}$ and  $\mathbf{A}_{p}^{-1}$ exists.
\end{assumption}

\subsection{The dynamics in the natural coordinates are positive and nonlinear}
\begin{defn}\label{defn: nonlinear dynamics}
	Defining the state $\mathbf{x}\in\mathbb{R}_{0}^{+}$ as the vector of species concentrations, the mass action kinetics of the constructed CRN result in
	\begin{eqnarray}
	&&\dot{\mathbf{x}}=
	\left(\mathbf{A}^{\boxbslash}-\mathbf{D}\left\{ \left|\mathbf{a}\right|\right\}\right)\mathbf{x}
	+\mathbf{Br} -\eta\left(\mathbf{Px}\right)\circ\mathbf{x}\label{eq: non-linear model a + A}
	\end{eqnarray}
	where $\mathbf{A}=\mathbf{A}^{\boxbslash}-\mathbf{D}\left\{ \left|\mathbf{a}\right|\right\}$ and
		%
	%
	\begin{eqnarray}
	&&\mathbf{x} 
	=\left[\begin{array}{c|c}
	\left(\mathbf{x}^{+}\right)^{T} & \left(\mathbf{x}^{-}\right)^{T}\end{array}\right]^{T}
	=\left[\begin{array}{ccc|ccc}
	x_{1}^{+}  & \ldots & x_{N}^{+} & x_{1}^{-}  & \ldots & x_{N}^{-}\end{array}\right]^{T}
	\label{eq: state partition}
	\\
	&&\mathbf{P}=\left[\begin{array}{cc}
	\mathbf{0} & \mathbf{I}\\
	\mathbf{I} & \mathbf{0}
	\end{array}\right]\Rightarrow\left(\mathbf{Px}\right)\circ\mathbf{x}=\left[\begin{array}{c}
	\mathbf{x}^{+}\circ\mathbf{x}^{-}\\
	\mathbf{x}^{+}\circ\mathbf{x}^{-}
	\end{array}\right]\label{eq: nonlinear flux}
	\end{eqnarray}
\end{defn}
Compared to~\eqref{eq: I/O system state space}, the model in~\eqref{eq: non-linear model a + A} includes the components from the bimolecular reactions $-\eta \mathbf{x}^{+}\circ\mathbf{x}^{-}$. The unimolecular reactions depend linearly on the state with $\mathbf{Ax}$, where by construction the catalysis rates end up on the off-diagonal elements $\mathbf{A}^{\boxbslash}\geq0$ and the degradation rates result in non-positive elements in the diagonal of $\mathbf{D}\left\{\mathbf{a}\right\}$ ($\mathbf{a}\leq0$). 
$\mathbf{A}$ is Metzler, and we can decompose the dynamics into non-negative and non-positive contributions where
$\mathbf{D}\left\{\mathbf{a}\right\}-\eta\left(\mathbf{Px}\right)\circ\mathbf{x}\leq0$, and $\mathbf{A}^{\boxbslash}\mathbf{x}+\mathbf{Br}\geq0$.
With $\mathbf{B}\geq0$, $\mathbf{r}\geq0$, and $\mathbf{g}\left\{ \mathbf{x}\right\} =-\eta\left(\mathbf{Px}\right)$, the following Lemma~\ref{lm: positivity} shows that the nonlinear dynamics in their natural coordinates in~\eqref{eq: non-linear model a + A} are non-negative.
\begin{lemma}\label{lm: positivity}
	For a vector function $\mathbf{g}\left\{ \mathbf{x}\right\}$, if $\mathbf{M}\in\mathcal{M}$, $\mathbf{v}\geq0$,
	and $\mathbf{x}\left(0\right)>0$, the dynamics $\dot{\mathbf{x}}=\mathbf{Mx}+\mathbf{x}\circ\mathbf{g}\left\{ \mathbf{x}\right\} +\mathbf{v}$ are non-negative.
\end{lemma}
\begin{pf}
	For each component $\dot{x}_{j}=\left[\mathbf{Mx}\right]_{j}+x_{j}\left[\mathbf{g}\left\{ \mathbf{x}\right\} \right]_{j}+v_j$. If $x_{j}=0$ and $\exists_{i\neq j}:x_i>0$, then $\dot{x}_{j}=\left[\mathbf{Mx}\right]_{j}+\mathbf{v}\geq0$ and the trajectory remains in $\mathbb{R}^{+}_0$. \qed
\end{pf}
Rewriting~\eqref{eq: non-linear model a + A} according to the partition in (\ref{eq: state partition}) so that 
\begin{eqnarray}
\begin{array}{c}
\dot{\mathbf{x}}^{+}=\mathbf{A}_{1}^{+}\mathbf{x}^{+}+\mathbf{A}_{2}^{-}\mathbf{x}^{-}+\mathbf{B}_{1}^{+}r^{+}-\eta\mathbf{x}^{+}\circ\mathbf{x}^{-}\\
\dot{\mathbf{x}}^{-}=\mathbf{A}_{2}^{+}\mathbf{x}^{+}+\mathbf{A}_{1}^{-}\mathbf{x}^{-}+\mathbf{B}_{1}^{-}r^{-}-\eta\mathbf{x}^{+}\circ\mathbf{x}^{-}
\end{array}\label{eq: decomposed nlin dyn}
\end{eqnarray}
we have matrices $\mathbf{A}$ and $\mathbf{B}$ structured into
\begin{subequations}\label{eq: matrix decomposed nonlinear model}
\begin{eqnarray}
&&\mathbf{A}=\left[\begin{array}{cc}
\mathbf{A}_{1}^{+} & \mathbf{A}_{2}^{-}\\
\mathbf{A}_{2}^{+} & \mathbf{A}_{1}^{-}
\end{array}\right],\: 
\mathbf{B}=\left[\begin{array}{cc}
\mathbf{B}_{1}^{+} & \mathbf{0}\\
\mathbf{0} & \mathbf{B}_{1}^{-}
\end{array}\right] \\
&& \mathbf{A}_{1}^{\pm}=\left(\mathbf{A}_{1}^{\pm}\right)^{\boxbslash}+\mathbf{D}\left\{ \mathbf{a}_{1}^{\pm}\right\} ,
\;\mathbf{a}_{1}^{\pm}\leq0,
\; \mathbf{A}_{2}^{\pm}=\left(\mathbf{A}_{2}^{\pm}\right)^{\boxbslash}
\end{eqnarray}
\end{subequations}
$\mathbf{A}_{j}^{\pm}\in \mathcal{M}$,  and from Definition~\ref{defn: nonlinear dynamics} the degradation rates are in the diagonal of $\mathbf{A}_1^\pm$, while $\mathbf{A}_2^\pm$ contain only the catalysis reactions used to represent subtraction.
Because the catalysis and degradation reactions are duplicated, both matrices $\mathbf{A}_{i}^{\pm}$ retain the same structure, but not necessarily the same parameterisation
(similarly for the pair $\mathbf{B}_{1}^{\pm}$). Matrices $\mathbf{A}_{i}^{+}$ and $\mathbf{B}_{1}^{+}$ are populated with the reaction rates $\gamma_{j}^{+}$, and their counterparts $\mathbf{A}_{i}^{-}$ and $\mathbf{B}_{1}^{-}$ with $\gamma_{i}^{-}$.

\begin{figure}[t]
	\begin{centering}
		\includegraphics[width=0.8\columnwidth]{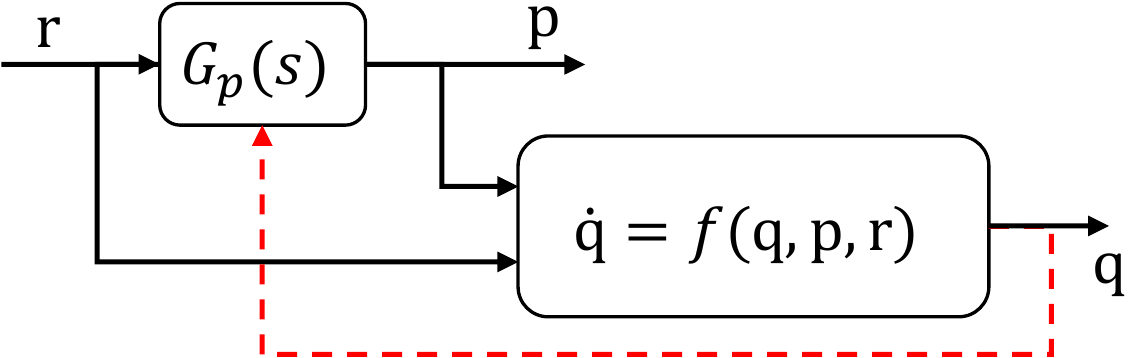}
		\par\end{centering}
	\caption{Interconnection between the I/O dynamics and the underlying positive dynamics in the rotated coordinates. The dashed connection is absent with the nominal symmetric parameterisation from Definition~\ref{defn: symmetrical matrices}.
	}
	\label{fig: cascade}
\end{figure}

\subsection{The positive nonlinear dynamics are unobservable in the I/O dynamics of the linear representation}

\begin{defn}\label{defn: rotated coordinates}
	The rotated coordinates $p_{j}=x_{j}^{+}-x_{j}^{-} \in\mathbb{R}$ and $q_{j}=x_{j}^{+}+x_{j}^{-} \in\mathbb{R}_0^+$ result from the similarity transformation $\mathbf{W}$, where
	\begin{eqnarray}
	&&\left[\begin{array}{c}
	\mathbf{p}\\
	\mathbf{q}
	\end{array}\right]
	=\left[\begin{array}{cc}
	\mathbf{I} & -\mathbf{I}\\
	\hdashline \mathbf{I} & \;\;\:\mathbf{I}
	\end{array}\right]\mathbf{x}
	=\left[\begin{array}{c}
	\mathbf{W}_{p}\\
	\hdashline \mathbf{W}_{q}
	\end{array}\right]\mathbf{x}	
	=\mathbf{W}\mathbf{x}\label{eq: Psi map}
	\end{eqnarray}
\end{defn}
We then have that $\mathbf{W}^{-1}=\frac{1}{2}\mathbf{W}^{T}$, $\mathbf{W}_{p}\left(\left(\mathbf{Px}\right)\circ\mathbf{x}\right)=0$ and $\mathbf{W}_{q}\left(\left(\mathbf{Px}\right)\circ\mathbf{x}\right)=2\eta\left(\mathbf{x}^{+}\circ\mathbf{x}^{-}\right)$.
%
%
We can use $\mathbf{W}$ to split the dynamics into the I/O dynamics $\dot{ \mathbf{p}}$ from Section~\ref{sec: dual rail CRN} and the remaining nonlinear positive dynamics $\dot{ \mathbf{q}}$, and infer their interconnections.
%
%
The rotated dynamics are then given by
	\begin{eqnarray}
	\begin{aligned}
	\left[\begin{array}{c}
	\dot{\mathbf{p}}\\
	\dot{\mathbf{q}}
	\end{array}\right]
	=&\left[\begin{array}{cc}
	\mathbf{R}_{11} & \mathbf{R}_{12}\\
	\mathbf{R}_{21} & \mathbf{R}_{22}
	\end{array}\right]\left[\begin{array}{c}
	\mathbf{p}\\
	\mathbf{q}
	\end{array}\right]+\left[\begin{array}{c}
	\mathbf{W}_{p}\\
	\mathbf{W}_{q}
	\end{array}\right]\mathbf{B}\mathbf{r}\\
	&-\frac{\eta}{2}\left[\begin{array}{c}
	0\\
	\mathbf{q}\circ\mathbf{q}-\mathbf{p}\circ\mathbf{p}
	\end{array}\right]
	\end{aligned}\label{eq: rotated non linear dynamics}
	\end{eqnarray}
\begin{remark}\label{rem: diagonal R22} 
From the structures in Definition~\ref{defn: nonlinear dynamics} and~\eqref{eq: matrix decomposed nonlinear model} (recall that $\mathbf{a}_{1}^{\pm}\leq0$), we have that
\begin{eqnarray}
	&&\begin{aligned}
	&\mathbf{R_{22}}=
	\frac{\left(\mathbf{A_{1}^{+}+A_{1}^{-}+A_{2}^{+}+A_{2}^{-}}\right)^{\boxbslash}}{2}
	-\frac{\mathbf{D}\left\{ \left|\mathbf{a_{1}^{+}}\right|+\left|\mathbf{a_{1}^{-}}\right|\right\}}{2}\\
	&\mathbf{R_{11}}=
	\frac{\left(\mathbf{A_{1}^{+}+A_{1}^{-}-A_{2}^{+}-A_{2}^{-}}\right)^{\boxbslash}}{2}
	-\frac{\mathbf{D}\left\{ \left|\mathbf{a_{1}^{+}}\right|+\left|\mathbf{a_{1}^{-}}\right|\right\}}{2}\\
	&\mathbf{R_{12}}=
	\frac{\left(\mathbf{A_{1}^{+}-A_{1}^{-}-A_{2}^{+}+A_{2}^{-}}\right)^{\boxbslash}}{2}
	-\frac{\mathbf{D}\left\{ \left|\mathbf{a_{1}^{+}}\right|-\left|\mathbf{a_{1}^{-}}\right|\right\}}{2} \\
	&\mathbf{R_{21}}=
	\frac{\left(\mathbf{A_{1}^{+}-A_{1}^{-}+A_{2}^{+}-A_{2}^{-}}\right)^{\boxbslash}}{2}
	-\frac{\mathbf{D}\left\{ \left|\mathbf{a_{1}^{+}}\right|-\left|\mathbf{a_{1}^{-}}\right|\right\}}{2} 
	\end{aligned}\nonumber
\end{eqnarray}
	The diagonal of ${\mathbf{R}}_{22}$ is nonpositive, given by the average of the diagonals of $\mathbf{A}^\pm_{1}$. Also $\mathbf{A}_j^\pm \in \mathcal{M}\Rightarrow \mathbf{R_{22}}\in\mathcal{M}$.
\end{remark}
\begin{defn}\label{defn: symmetrical matrices}
	%
	Considering the condition of perfectly identical reaction rates from Assumption~\ref{assum: rates are equal}, we define the nominal  matrices (represented with an upper bar), where we have that
	$\mathbf{A}_{1}^{\pm}=\bar{\mathbf{A}}_{1}$, $\mathbf{A}_{2}^{\pm}=\bar{\mathbf{A}}_{2}$, $\mathbf{B}_{1}^{\pm}=\bar{\mathbf{B}}_{1}$.
\end{defn}
\begin{proposition}\label{lem: unobservable nominal dynamics}
	For the nominal symmetrical parameterisation in Definition~\ref{defn: symmetrical matrices}, the nonlinear dynamics are unobservable in the I/O system, due to the serial structure of the nominal rotated dynamics given by
	\begin{subequations}\label{eq: rotated nominal}
	\begin{eqnarray}
	\dot{\mathbf{p}}&=&  \bar{\mathbf{R}}_{11}\mathbf{p}+\mathbf{W}_{p}\bar{\mathbf{B}}\mathbf{r}\label{eq: rotated nominal linear model}\\
	\dot{\mathbf{q}}&=&  \bar{\mathbf{R}}_{22}\mathbf{p}+\mathbf{W}_{q}\bar{\mathbf{B}}\mathbf{r}+\frac{\eta}{2}\mathbf{p}\circ\mathbf{p}-\frac{\eta}{2}\mathbf{q}\circ\mathbf{q}
	\label{eq: rotated nominal nonlinear model}
	\end{eqnarray}
	\end{subequations}
\end{proposition}
\begin{pf} 
	Applying Definition~\ref{defn: symmetrical matrices} to the matrices in Remark~\ref{rem: diagonal R22}, it follows immediately that $\bar{\mathbf{R}}_{12}=\bar{\mathbf{R}}_{21}=0$, $\bar{\mathbf{R}}_{11}=\bar{\mathbf{A}}_{1}-\bar{\mathbf{A}}_{2}$, $\bar{\mathbf{R}}_{22}=\bar{\mathbf{A}}_{1}+\bar{\mathbf{A}}_{2}$, and thus the serial structure of~(\ref{eq: rotated nominal linear model}-\ref{eq: rotated nominal nonlinear model}) (illustrated in Fig.~\ref{fig: cascade}) means that $\mathbf{p}$ evolves independently of $\mathbf{q}$, making $\mathbf{q}$ unobservable in any output of the I/O dynamics.
	\qed
	%
\end{pf}
%

%
%
\section{Equilibria of the chemical reaction network }
We now compare the equilibria of the CRN with and without feedback, to analyse how feedback changes the fundamental properties of the system.
%
%
\begin{defn}\label{defn: cascaded system}
	We define a cascaded system as a set of DSD reactions without feedback, where the catalysis reactions do not depend directly or indirectly on the chemical species downstream.
\end{defn}
Cascaded strand displacement reactions are well suited to systematically build large computational and logic gate circuitry~\cite{Qian2011c}. 
The cascaded structure of the represented linear system results in a state matrix which can be permuted so that $\bar{\mathbf{R}}_{11}=\mathbf{A}_{p}\in\mathcal{L}$. Under Assumptions~\ref{assum: rates are equal} and~\ref{R11 stable}, and from Remark~\ref{rem: diagonal R22}, we have $\bar{\mathbf{R}}_{11}\in\mathcal{L}\Rightarrow \bar{\mathbf{A}}_{1},\bar{\mathbf{A}}_{2}\in\mathcal{L}$, and $\bar{\mathbf{R}}_{11}\in\mathcal{L,H}\Rightarrow \bar{\mathbf{R}}_{22}\in\mathcal{L,H}$. 

\begin{figure}[t]
	\begin{centering}
		\includegraphics[width=1\columnwidth]{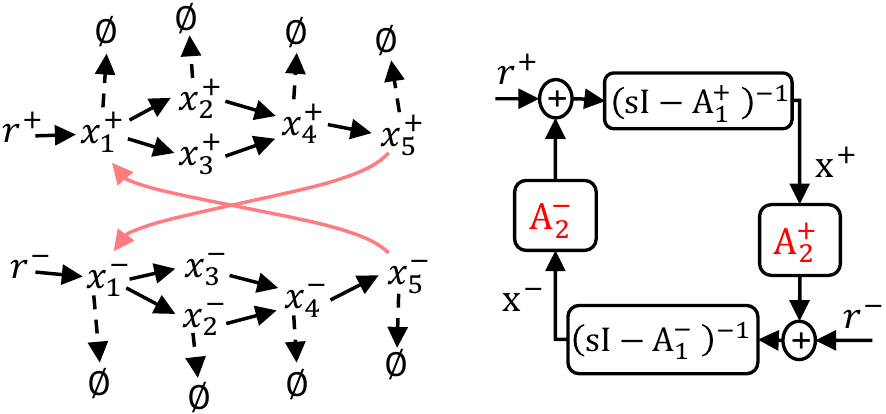} \par 
		\hspace{0.25\columnwidth}\textbf{a)}  \hspace{0.45\columnwidth} \textbf{b)}  
	\end{centering}
	\caption{\textbf{a)} Network of catalysis (full line) and degradation (dashed) reactions for Example~\ref{exmp: ODEs CRN}. Without the negative feedback the system is a cascade of reactions (black arrows) from inputs $r^\pm$ to the outputs $x_5^\pm$ and $\mathbf{A}_{2}^\pm=0$. \textbf{b)} Introducing the negative feedback from $\mathbf{A}_{2}^{\pm}\geq0$ (red arrows), introduces positive feedback between positive systems.}
	\label{fig_metzler_part}
\end{figure}

For example, representing the open loop of Fig.~\ref{fig: Feedback-interconnection-linear} without the reactions $X_{5}^{\pm}\xrightarrow{\gamma_{2}}X_{5}^{\pm}+X_{1}^{\mp}$ results in the cascade of serial and parallel reactions in Fig.~\ref{fig_metzler_part}a.
In this particular case it also results in $\bar{\mathbf{A}}_{2}=0$, but in general, we can have $\bar{\mathbf{A}}_{2}\geq 0$ if there are subtractions in the cascaded I/O dynamics.
Including feedback in the I/O dynamics leads to feedback within the network, and the cascaded structure is lost. The reactions
$X_{5}^{\pm}\xrightarrow{\gamma_{2}}X_{5}^{\pm}+X_{1}^{\mp}$ connect the output to the input of the open loop cascade of reactions, and mass is transferred back into the input of the CRN.

Due to the triangular structure, the equilibrium of the unforced dynamics can be easily computed sequentially for each coordinate to show that there is a unique equilibrium at $\mathbf{q}=0$ \commentarxiv{(Lemma~\ref{lem: equilibrium L nominal} in Appendix)}. In the presence of feedback this is no longer possible since the states will depend on the output, and it follows that  $\exists_{i>j}:\left[{\mathbf{A}}^\pm_2 \right]_{ji}>0$, and $\bar{\mathbf{A}}_2\notin\mathcal{L}$. 
Consequently, the states involved in the closed loop become interdependent, and $\bar{\mathbf{R}}_{11}=\bar{\mathbf{A}}_{1}-\bar{\mathbf{A}}_{2}$ cannot be a lower triangular matrix.
\begin{remark}\label{lem: irreducible dyn}
	The interdependent evolution of all the states is reflected in the \emph{irreducibility} \cite{Farina2000} of the state matrix $\bar{\mathbf{R}}_{22}$.
	If $\bar{\mathbf{R}}_{22}\in \mathcal{I},\mathcal{M}$, for each coordinate
	$j$, $\exists_{i}:\left[\bar{\mathbf{R}}_{22}\right]_{ji}>0$. Therefore the trajectory of $q_{j}$ will always depend on another coordinate $q_{i}$, making the network irreducible.
\end{remark}
\begin{proposition}\label{thm: general solution of eq.}
	Consider $\mathbf{M}\in \mathcal{I},\mathcal{M}$ such that
	$\mathbf{M}=\mathbf{M}^{\boxbslash}+\mathbf{D}\left\{ \mathbf{m}\right\}$, $\mathbf{m}\leq 0$, and the dynamics $\dot{\mathbf{q}}=\mathbf{M}\mathbf{q}-k\mathbf{q}\circ\mathbf{q}$ with equilibrium $\mathbf{q}^*$. Then we have the following: i) $\exists_j q_j^*=0\Rightarrow q_{i\neq j}^*=0$; ii) the unforced dynamics may
	admit a second positive equilibrium $\mathbf{q}^*>0$, proportional to $k^{-1}$.
\end{proposition}
\begin{pf}
	From the equilibrium condition for each coordinate $j$ we take the non-negative roots
	\begin{eqnarray}
	&&kq_{j}^{2}+|m_{jj}|q_{j}-\sum_{i\neq j}m_{ji}q_{i}=0\\
	&&\Rightarrow q_{j}=\frac{1}{2k}\left(-|m_{jj}|+\sqrt{m_{jj}^{2}+4k\sum_{i\neq j}m_{ji}q_{i}}\right)\geq 0\label{eq: roots feedback eq}
	\end{eqnarray}
	\textbf{i)} If $\sum_{i\neq j}m_{ji}q_{i}=0$, then $q_{j}=0$, and we disregard
	the negative solution $q_{j}=-|m_{jj}|/k$. Since
	$\mathbf{M}\in\mathcal{I}$, for every coordinate $j$, $\exists_{l\neq j}:m_{jl}>0$, and $q_{j}=0\Leftrightarrow q_{l}=0$.
	We also have that for any $i\neq j:m_{ij}>0$,
	$q_{i}=0\Leftrightarrow q_{j}=0$. Hence, if $q_{j}=0\Rightarrow\forall_{i\neq j},\:q_{i}=0$, and we cannot have an equilibrium where only some of the states are at zero. 
	
	\textbf{ii)} If $\exists_{i\neq j}:m_{ji}>0$ and the coordinate $i$ is at a positive equilibrium $q_{i\neq j}^{*}>0$, then $\sum_{i\neq j}m_{ji}q_{i}^{*}>0$. The non-negative roots for each coordinate $j$ result from solving the system~\eqref{eq: roots feedback eq}. Note that even if $m_{jj} =0$, then $q_{j}^{*}>0$. 

	Combining \textbf{i)} and \textbf{ii)}, if $\mathbf{M}\in\mathcal{I}$, the system may have a positive equilibrium $\mathbf{q}^{*}>0$, which can be scaled down with $k$, since $\lim_{k\rightarrow\infty}q_{j}^{*}=0$.\qed
\end{pf}
\begin{example}\label{exmp: 2 state example} 
Consider the CRN representation of a linear system with a single input $u$, which has negative feedback between its states $x$ and $y$ ($c_2>0$), resulting in
	\begin{eqnarray}
	&&\dot{x}  =-d_{1}x-\boldsymbol{c_{2}y}+u, \quad \dot{y}  =-d_{2}y+c_{1}x
	\\&&\Rightarrow \begin{cases}
	\begin{aligned}
	U^{\pm}\xrightarrow{1}U^{\pm}+X^{\pm},\:X^{\pm}\xrightarrow{d_{1}}\emptyset,\:X^{+}+X^{-}\xrightarrow{k}\emptyset\\
	X^{\pm}\xrightarrow{c_{1}}X^{\pm}+Y^{\pm},\:Y^{\pm}\xrightarrow{d_{2}}\emptyset,\:Y^{+}+Y^{-}\xrightarrow{k}\emptyset\\
	Y^{\pm}\xrightarrow{c_{2}}Y^{\pm}+X^{\mp},\:U^{+}+U^{-}\xrightarrow{k}\emptyset
	\end{aligned}\label{crn: example 2}	\end{cases}\\
&&\Rightarrow \bar{\mathbf{R}}_{11}=\left[\begin{array}{cc}
-d_{1} & -c_{2}\\
c_{1} & -d_{2}
\end{array}\right],\quad 
\bar{\mathbf{R}}_{22}=\left[\begin{array}{cc}
-d_{1} & c_{2}\\
c_{1} & -d_{2}
\end{array}\right]\label{eq: example 2 R matrices}
\end{eqnarray}
\end{example}
Without feedback, so that $c_{2}=0$, the system simplifies to a reducible serial cascade where $\bar{\mathbf{R}}_{11}=\bar{\mathbf{R}}_{22}=\bar{\mathbf{A}}_{1}\in\mathcal{L}$, and the unforced dynamics $\dot{\mathbf{q}}=\bar{\mathbf{R}}_{22}\mathbf{q}-k\mathbf{q}\circ\mathbf{q}$ have a single non-negative equilibrium at $\mathbf{q}=0$.
%

%
With feedback, so that $c_{2}>0$, we can replace $q_{2}=c_{2}^{-1}\left(kq_{1}+d_{1}\right)q_{1}$ in the equilibrium conditions for $q_1$ and obtain the polynomial
\begin{eqnarray}
\begin{aligned}
k^{3}q_{1}^{4}+2k^{2}d_{1}q_{1}^{3}&+\left(d_{1}^{2}+c_{2}d_{2}\right)kq_{1}^2& \\  
&+c_{2}\left(d_{2}d_{1}-c_{2}c_{1}\right)q_1=0&
\end{aligned}
\end{eqnarray}
Using Descartes' rule of signs, if $c_2>{d_2 d_1}{c_1^{-1}}$, we have one positive root and the equilibrium $q_1^*>0$ exists.
%
\begin{remark}\label{prop: negative vs positive feedback}
	Note that the use of $\bar{\mathbf{A}}_2$ to represent negative feedback in the I/O dynamics~(\ref{eq: I/O system state space}) with ${\mathbf{A}}_{p}=\bar{\mathbf{R}}_{11}=\bar{\mathbf{A}}_1-\bar{\mathbf{A}}_2$, results in positive feedback in the nonlinear dynamics in~\eqref{eq: rotated nominal nonlinear model} with $\bar{\mathbf{R}}_{22}=\bar{\mathbf{A}}_1 + \bar{\mathbf{A}}_2$
\end{remark}
In~\eqref{eq: example 2 R matrices}, 
$c_2$ impacts  the spectral radius of $\bar{\mathbf{R}}_{11}$ and $\bar{\mathbf{R}}_{22}$ differently. From their characteristic polynomials,  we have stable I/O dynamics ($\bar{\mathbf{R}}_{11}\in\mathcal{H}$) for any $c_2>0$, but for a sufficiently high gain $ c_{2}>{d_2 d_1}{c_1^{-1}}$, we get $\bar{\mathbf{R}}_{22}\notin\mathcal{H}$. Not coincidentally, it is the same domain for which $\mathbf{q}^*>0$ exists. \commentarxiv{Further details in the Appendix}.%
\begin{remark}\label{rem: experimental q>0}
	The existence of positive equilibrium conditions for linear feedback systems has direct consequences for the experimental construction of these circuits. 
	Operating at an equilibrium corresponding to high concentrations aggravates leaky reactions, where undesired triggering of strand displacement leads to unwanted outputs in the absence of inputs.
	Furthermore, if ${\mathbf{q}}^*\geq0$ with input $\mathbf{r}=0$, then the reactions persist even if the I/O dynamics are at rest $\mathbf{p}=0$, leading to unnecessary, irreversible, and costly consumption of fuel species. 
	%
	%
	%
	This is in direct contrast to cascaded DSD reactions, where without input to the I/O dynamics, the CRN is at equilibrium at $\mathbf{x}=0$, and no reactions occur. 	
\end{remark}

\myclearpage

%
%
\section{Stability }

We begin by proving the following lemma which is applicable to the unforced dynamics of~\eqref{eq: non-linear model a + A} and~\eqref{eq: rotated nominal nonlinear model}. 
\begin{lemma}\label{lm: stable if M hurwitz}
	If $\mathbf{M}\in\mathcal{M},\mathcal{H}$, and $\mathbf{g}\left\{ \mathbf{x}\right\} <0$ for $ \mathbf{x}>0$, then the system $\dot{\mathbf{x}}=\mathbf{Mx}+\mathbf{x}\circ\mathbf{g}\left\{ \mathbf{x}\right\} $ is globally asymptotically stable (GAS) at $\mathbf{x}=0$.
\end{lemma} 
\begin{pf} 
From the stability of Metzler matrices~\cite{Farina2000}, $\mathbf{M}\in\mathcal{M,H}\Rightarrow\exists_{\mathbf{d}>0}:\mathbf{M}^{T}\mathbf{D}\{\mathbf{d}\}+\mathbf{D}\{\mathbf{d}\}\mathbf{M}=-\mathbf{I}$.
	We take the Lyapunov function $V_d\left\{ \mathbf{x}\right\} =\mathbf{x}^{T}\mathbf{D}\left\{ \mathbf{d}\right\} \mathbf{x}>0$, and since $\mathbf{D\{d\}}\left(\mathbf{x}\circ\mathbf{g}\left\{ \mathbf{x}\right\} \right)=\mathbf{d}\circ\mathbf{x}\circ\mathbf{g}\left\{ \mathbf{x}\right\}<0 $, $\forall_{\mathbf{x}>0}$, we have that
		$\dot{V}_d\left(\mathbf{x}\right)
= -\mathbf{I}+2\mathbf{g}\left\{ \mathbf{x}\right\} ^{T}\left(\mathbf{d}\circ\mathbf{x}\circ\mathbf{x}\right)<0 $\qed
\end{pf}
With $\mathbf{g}\left\{ \mathbf{x}\right\} =-\mathbf{Px}$, Lemma~\ref{lm: stable if M hurwitz} ensures that if the network of catalysis and degradation reactions is stable, $\mathbf{A}\in\mathcal{H}$, the bimolecular reactions cannot destabilise~\eqref{eq: non-linear model a + A}.
A stable CRN with $\mathbf{A}\in\mathcal{H}$ can occur if the degradation of each species is faster than their overall production, and $\mathbf{A}$ has a dominant diagonal. However, this is not the general case.
The dynamics without the bimolecular reactions result in the positive feedback loop between two positive systems of Fig.~\ref{fig_metzler_part}b.
Since we cannot stabilise non-negative systems with positive gains $\mathbf{A}_2^\pm\geq 0$~\cite{Roszak2009}, it is sufficient to have $\mathbf{A}_{1}^{\pm}\notin\mathcal{H}$ to give $\mathbf{A}\notin\mathcal{H}$.
Even for the nominal symmetrical parameterisation, the representation has modes that are not present in the original linear system $\rho\left\{\bar{\mathbf{R}}\right\}=\rho\left\{\bar{\mathbf{R}}_{11}\right\}\cup\rho\left\{\bar{\mathbf{R}}_{22}\right\}$. 
While this is a problem for IPR with linear positive systems~\cite{Cacace2012}, the presence of the bimolecular reactions are sometimes sufficient for stabilisation, even if  $\bar{\mathbf{R}}_{22}\notin\mathcal{H}$.

 %

%
%
\subsection{The I/O dynamics determine the stability for the nominal symmetrical case}

While at first glance it seems precarious to have unobservable nonlinear dynamics, for the nominal symmetrical case in Definition~\ref{defn: symmetrical matrices}, it is possible to provide guarantees for stability and boundedness.
%
%
\begin{proposition}\label{thm: GAS cascaded}
	The cascaded systems from Definition~\ref{defn: cascaded system} representing stable I/O dynamics, have GAS unforced nonlinear dynamics, for $\mathbf{x}>0$.
\end{proposition}
\begin{pf} 
	From Remark~\ref{rem: diagonal R22}, in cascaded systems  $\bar{\mathbf{R}}_{11},\bar{\mathbf{R}}_{22}\in\mathcal{L}$, and $ \rho\left\{\bar{\mathbf{R}}_{11}\right\}=\rho\left\{\bar{\mathbf{R}}_{22}\right\}$.
	If the I/O system is stable, then $\alpha\left\{\bar{\mathbf{R}}_{11}\right\}=\alpha\left\{\bar{\mathbf{R}}_{22}\right\}<0$ and Lemma~\ref{lm: stable if M hurwitz} ensures $\mathbf{\dot{q}}=\mathbf{\bar{R}_{22}}\mathbf{q}-\frac{\eta}{2}\mathbf{q}\circ\mathbf{q}$
	is GAS at $\mathbf{q}= 0$.\qed
\end{pf}
%
%
%
\begin{remark}
	We can apply Proposition~\ref{thm: GAS cascaded} to the representation of individual linear operations, which by themselves are cascaded reactions. It results directly that the CRNs for summation, gain, and subtraction by themselves, have GAS unforced dynamics, and are bounded for bounded inputs.
	More importantly, applying it to CRNs assembled from those linear operations in a cascade fashion, results in a single stable equilibrium for the complete circuit.
\end{remark}  
%
%
%
Recalling that with the introduction of feedback, we lose the cascaded structure and create an irreducible  system, even for the representation of stable I/O linear dynamics ($\bar{\mathbf{R}}_{11}\in\mathcal{H}$), if feedback leads to $\bar{\mathbf{R}}_{22}\notin\mathcal{H}$, then the following Lemma states that unforced trajectories diverge away from the origin due to a diverging mode of $\bar{\mathbf{R}}_{22}$.
\begin{lemma}
	For the dynamics $\dot{\mathbf{q}}={\mathbf{M}}\mathbf{q}-k\mathbf{q}\circ\mathbf{q}$ with ${\mathbf{M}}\in\mathcal{M,I}$ but ${\mathbf{M}}\notin\mathcal{H}$, the equilibrium at the origin $\mathbf{q}=0$ is unstable.
	\label{lem: unstable nom q=0}
\end{lemma}
\begin{pf} 
From applying the Frobenius-Perron theorem to Metzler matrices~\cite{Farina2000,Horn2012matrix},  $\mathbf{M}\in\mathcal{M,I}\Rightarrow\exists_{\mathbf{w_F>0}}: \mathbf{w}_{F}^{T}{\mathbf{M}}=\lambda_{F}\mathbf{w}_{F}^{T}$ and $\lambda_F =\alpha\left\{\mathbf{M}\right\}$.
	%
	Defining the Lyapunov function $V_F\left(\mathbf{q}\right)=\mathbf{w}_{F}^{T}\mathbf{q}$, we have that $\mathbf{q}>0\Rightarrow V_F\left(\mathbf{q}\right)>0$ and
	$\dot{V}_F\left(\mathbf{q}\right) =\mathbf{w}_{F}^{T}\dot{ \mathbf{q}}  
	=\mathbf{w}_{F}^{T}\left(\mathbf{q}\circ\left(\lambda_{F}\mathbf{1}-k\mathbf{q}\right)\right)$.
	%
	%
	Since ${\mathbf{M}}\notin\mathcal{H}\Rightarrow\lambda_{F}>0$, hence $\forall_{j},\,q_{j}<\frac{\lambda_{F}}{k}$ gives that $\dot{V}_F\left\{ \mathbf{q}\right\} >0$,  and the system is divergent close to the origin.
	\qed
\end{pf} 
The IPR of a stable system using only linear positive systems is therefore not guaranteed to be stable~\cite{Cacace2012}. However, for the nonlinear positive dynamics~\eqref{eq: rotated nominal}, we can still ensure boundedness with the following result.
\begin{lemma}\label{lem: bounded nominal q}
	For $\mathbf{M}\in\mathcal{M}$ , $\mathbf{q}\left(0\right)>0$,
	and a bounded input $\mathbf{v}\geq0$, if $\mathbf{g}\left\{ \mathbf{q}\right\} \leq-k\mathbf{q}$
	then the non-negative trajectories of $\dot{\mathbf{q}}=\mathbf{M}\mathbf{q}+\mathbf{q}\circ\mathbf{g}\left\{ \mathbf{q}\right\} +\mathbf{v}$
	are bounded by $\|\mathbf{q}\|_{2}<k^{-1}\left(\sqrt{N}\|\mathbf{M}\|_{2}+\|\mathbf{v}\|_{1}\|\mathbf{q}\|_{2}^{-1}\right)$
\end{lemma}
\begin{pf} 
		Lemma~\ref{lm: positivity} guarantees that the trajectories are nonnegative for
	$\mathbf{q}\left(0\right)>0$. If $\mathbf{M}\in\mathcal{H}$ is Hurwitz, Lemma~\ref{lm: stable if M hurwitz} guarantees that the system is asymptotically stable in $\mathbb{R}_{0}^{+}$
	with equilibrium at $\mathbf{q}=0$. If $\mathbf{M}\notin\mathcal{H}$ is not Hurwitz,
	we can still show boundedness, using the linear Lyapunov function
	$V_1\left\{ \mathbf{q}\right\} =\|\mathbf{q}\|_1=\sum_{j}q_{j}>0$, in the domain $\mathbf{q}>0$. We then have
	\begin{eqnarray*}
		&\dot{V}_1\left\{ \mathbf{q}\right\}  & 
		  = \mathbf{1}^{T}\mathbf{M}\mathbf{q}+\mathbf{1}^{T}\mathbf{v}+\mathbf{1}^{T}\mathbf{D\left\{ \mathbf{q}\right\} }\mathbf{g}\left\{ \mathbf{q}\right\} \\
		 && = \mathbf{1}^{T}\mathbf{M}\mathbf{q}+\|\mathbf{v}\|_{1}+\mathbf{q}^{T}\mathbf{g}\left\{ \mathbf{q}\right\} \\
&&  		\leq \|\mathbf{M}\mathbf{q}\|_{1}+\|\mathbf{v}\|_{1}-k\mathbf{q}^{T}\mathbf{q}\\
		 && \leq \sqrt{N}\|\mathbf{M}\|_{2}\|\mathbf{q}\|_{2}+\|\mathbf{v}\|_{1}-k\|\mathbf{q}\|^2_{2}
	\end{eqnarray*}
	We can always find large enough values of $\mathbf{q}$ such that
	$ \|\mathbf{q}\|_{2}>\frac{\sqrt{N}}{k}\|\mathbf{M}\|_{2}+\frac{1}{k}\frac{\|\mathbf{v}\|_{1}}{\|\mathbf{q}\|_{2}}$
	where we have $\dot{V}_1\left\{ \mathbf{q}\right\} <0$. 
	\qed	
\end{pf}
Applying Lemma~\ref{lem: bounded nominal q} with $\mathbf{g}\left\{ \mathbf{q}\right\} =-\frac{\eta}{2}\mathbf{q}$ to the unforced dynamics in~\eqref{eq: rotated nominal nonlinear model} we have $\|\mathbf{q}\|_{2}<\eta^{-1}2\sqrt{N}\|\bar{\mathbf{R}}_{22}\|_{2}$ 
\commentarxiv{(see illustration in Fig. \ref{fig: q0 Vdot} in Appendix)}.
In general, Lemma~\ref{lem: bounded nominal q} is not applicable to the nonlinear dynamics (\ref{eq: non-linear model a + A}), due to the matrix $\mathbf{P}$. Moreover, it relies on the assumption of a stable $\bar{\mathbf{R}}_{11}$.
\begin{proposition}\label{prop: stability for nominal}
	Consider the nominal dynamics in~(\ref{eq: rotated nominal linear model}-\ref{eq: rotated nominal nonlinear model}), with the symmetrical parameterisation from Assumption~\ref{assum: rates are equal}. Under Assumption~\ref{R11 stable}, the I/O dynamics~\eqref{eq: rotated nominal linear model} are stable, and the concentrations in the complete CRN are bounded and can be scaled down with a faster annihilation reaction rate $\eta$.
\end{proposition}
\begin{pf}	
	Assumption~\ref{R11 stable} ensures the trajectories of $\mathbf{p}$ are bounded.
	%
	We can treat $\mathbf{p}$ as an additional input to the system~\eqref{eq: rotated nominal nonlinear model} and apply Lemma~\ref{lem: bounded nominal q} with $\mathbf{v}=\mathbf{W}_{q}\bar{\mathbf{B}}\mathbf{r}+\frac{\eta}{2}\mathbf{p}\circ\mathbf{p}$.
	The unobserved dynamics are then bounded for bounded inputs $\mathbf{r},\mathbf{p}>0$, and are scaled down by increasing $\eta$.
	%
\qed
\end{pf}
	The same feedback responsible for a stable I/O linear dynamics can result in $\bar{\mathbf{R}}_{22}\notin\mathcal{H}$ (see Remark~\ref{prop: negative vs positive feedback}).
	Designing feedback to ensure that $\bar{\mathbf{R}}_{11},\bar{\mathbf{R}}_{22}\in\mathcal{H}$ is impractical since it would put constraints on which I/O systems could be represented. It is one of the challenges of representing stable linear systems relying only on linear positive systems~\cite{Cacace2012}, where we would need $\bar{\mathbf{A}}\in\mathcal{H}$ for the IPR to be stable. Lemma~\ref{lem: bounded nominal q} lifts this constraint, albeit at the cost of a positive equilibrium.
\begin{remark}\label{thm: q>0 for irreducible}
	With the introduction of feedback,  the concentrations involved in the irreducible parts of the CRN will have positive equilibria, and $\exists_{j}\;q_{j}(t)>0$ even if $\mathbf{r}=0$ and the I/O dynamics are stable $\alpha\left\{ \bar{\mathbf{R}}_{11}\right\}<0$. This result also explains why in experimental practice the annihilation rate $\eta$ is set as high as possible, to minimise the concentrations in the circuit during operation or at equilibrium.
\end{remark}
\begin{remark}
	In the presence of integrators $\bar{\mathbf{a}}_1\leq 0$, it is not possible to use positive feedback $\bar{\mathbf{A}}_2$ such that $\bar{\mathbf{A}}_1+\bar{\mathbf{A}}_2$ becomes Hurwitz~\cite{DeLeenheer2001,Roszak2009}. 
	Starting from a  marginally stable state matrix $\alpha\left\{ \bar{\mathbf{A}}_{1}\right\} =0$,  the introduction of feedback leads to $\alpha\left\{\bar{\mathbf{R}}_{22}\right\}\geq 0$.
	This raises an interesting tradeoff, when controllers that introduce integrators in the loop transfer function (for example in PI control) lead to a positive equilibrium, which is inconvenient for implementation. 
\end{remark}

\subsection{Local stability with asymmetrical parameterisation from experimental variability}

The construction of the I/O dynamics in (\ref{eq: I/O system state space}) assumes the symmetrical parameterisation in Definition~\ref{defn: symmetrical matrices}. 
For a parametric analysis of the I/O system $\dot{\mathbf{p}}=\mathbf{A}_{p}\mathbf{p}+\mathbf{B}_{p}r$ Assumption~\ref{assum: rates are equal} still holds. 
Hence, as long as the I/O linear dynamics are stable, Proposition~\ref{prop: stability for nominal} guarantees that the nonlinear dynamics are bounded.

Once we (realistically) allow that all the parameters in Example~\ref{exmp: ODEs CRN} can vary independently, we get an asymmetric parameterisation that deviates from Assumption~\ref{assum: rates are equal}.
The dynamics for the I/O signals $p_{j}$ are still linear ($\mathbf{W}_{p}\left(\mathbf{Px}\circ\mathbf{x}\right)=0$), however, they depend on the nonlinear dynamics through the term $\mathbf{R}_{12}\mathbf{q}$  (absent in~\eqref{eq: I/O system state space} and~\eqref{eq: rotated nominal linear model})
\begin{eqnarray}
\dot{\mathbf{p}} & 
=&\mathbf{R}_{11}\mathbf{p}+\mathbf{W}_{p}\mathbf{B}\mathbf{r}+\mathbf{R}_{12}\mathbf{q}
\end{eqnarray}
\begin{remark}\label{rem: analysis of interconnection}
	With experimental variability, we lose the serial structure from (\ref{eq: rotated nominal nonlinear model}), and the I/O linear system and the underlying positive dynamics become interconnected (dashed connection in Fig. \ref{fig: cascade}).
	A stable I/O dynamics $\mathbf{R}_{11}\in\mathcal{H}$ no longer provides guarantees of boundedness, since it ignores the feedback between the I/O linear dynamics and the underlying nonlinear dynamics. Therefore, we need to analyse the stability of the complete nonlinear dynamics of (\ref{eq: non-linear model a + A}). 
\end{remark}
We investigate the stability of the nonlinear system using Lyapunov's indirect method, and the eigenvalues of the linearisation at the equilibrium of the system. 
For an equilibrium $\mathbf{x}=\mathbf{x}^{*}$, $\mathbf{r}=0$, and $
\mathbf{J}\left\{ \mathbf{x}^{*}\right\} =-\mathbf{D}\left\{ \mathbf{P}\mathbf{x}^{*}\right\} -\mathbf{D}\left\{ \mathbf{x}^{*}\right\} \mathbf{P}
$, the linearisation of (\ref{eq: non-linear model a + A}) results in the following
\begin{eqnarray}
\dot{\mathbf{s}} & =\left(\mathbf{A}+\eta\mathbf{J}\left\{ \mathbf{x}^{*}\right\} \right)\mathbf{s}+\mathbf{B}\mathbf{r}_{e}=\mathbf{A}_{s}\mathbf{s}+\mathbf{B}\mathbf{r}_{e}
\end{eqnarray}
If $\alpha\left\{ \mathbf{A}_s\right\} <0$ then the
system is locally exponentially stable around the equilibrium~\cite{Khalil2015}.
The equilibrium $\mathbf{x}^{*}=0$ is stable if and only if $\mathbf{A}\in\mathcal{H}$, in agreement with Lemma~\ref{lm: stable if M hurwitz}.
With the participation of $\mathbf{J}\left\{ \mathbf{x}^{*}\right\} $,	even if $\mathbf{A}$ is not Hurwitz, the linearisation can still	be stable around the equilibrium $\mathbf{x}^{*}>0$, showing the stabilising role of the bimolecular reactions.
It is also noteworthy that $\mathbf{W}_{p}\mathbf{J}\left\{ \mathbf{x}^{*}\right\} =0$, hence $\mathbf{A}_p$ and the stability of the linear I/O dynamics does not depend on the equilibrium. 

\commentarxiv{For the particular case of cascaded systems, as long as all species degrade with some non-zero rate, we show in Appendix that the CRN  is stable.}

%
%

\section{Analysis of an example nucleic acid feedback control system }
\begin{table}[t]
	\caption{ Nominal parameters for the example, and an asymmetrical parameterisation case which results in unstable dynamics. }
	\label{tab: parameters}
	\centering{}
	\begin{tabular}{|p{55pt}|p{50pt}|p{100pt}|}
		\hline
		Parameter & Nominal & Asymmetrical case \\
		\hline
		$k_{1}^{\pm}$ & $0.001/$s & $0.00132/$s\\ \hline
		$k_{2}^{\pm}$ & $0.001/$s & $k_2^+= 0.001320 $/s, \par $k_2^-=0.000680$/s\\ \hline
		$\gamma_{i}^{\pm}$, \par $i=1,2,3$ & $0.004/$s & $\gamma_1^\pm=\gamma_2^{\pm}=0.00528/$s, \par $\gamma_3^\pm=0.00272/$s  \\ \hline
		$\gamma_{i}^{\pm}$, \par $i=6,7,8$ & $0.008/$s & $\gamma_{6}^{\pm}=\gamma_{7}^{+}=0.01056/$s, \par $\gamma_{7}^-=\gamma_{8}^{\pm}=0.00544/$s \\ \hline
		$\gamma_{i}^{\pm}$, \par $i=4,5$ & $4\times10^{-6}/$s & $\gamma_4^+=2.72\times 10^{-6}/$s,  \par  $\gamma_4^-=\gamma_5^{\pm}=5.28\times 10^{-6}/$s\\ \hline
		$k_{P}^{\pm}=\gamma_{4}^\pm /\gamma_{5}^\pm$ & $1$ & $k_{P}^+=0.5152$, $k_{P}^-=1$\\ \hline
		$k_{0}^{\pm}$ & $0.00045/$s & $0.000594/$s \\ 
		\hline
		$\eta$ & $5\times10^{5}/\text{M}/$s & $5\times10^{5}/\text{M}/$s  \\ 		\hline
	\end{tabular}
	\label{tab1}
\end{table}

\begin{figure}[]
	\centerline{\includegraphics[width=1\columnwidth]{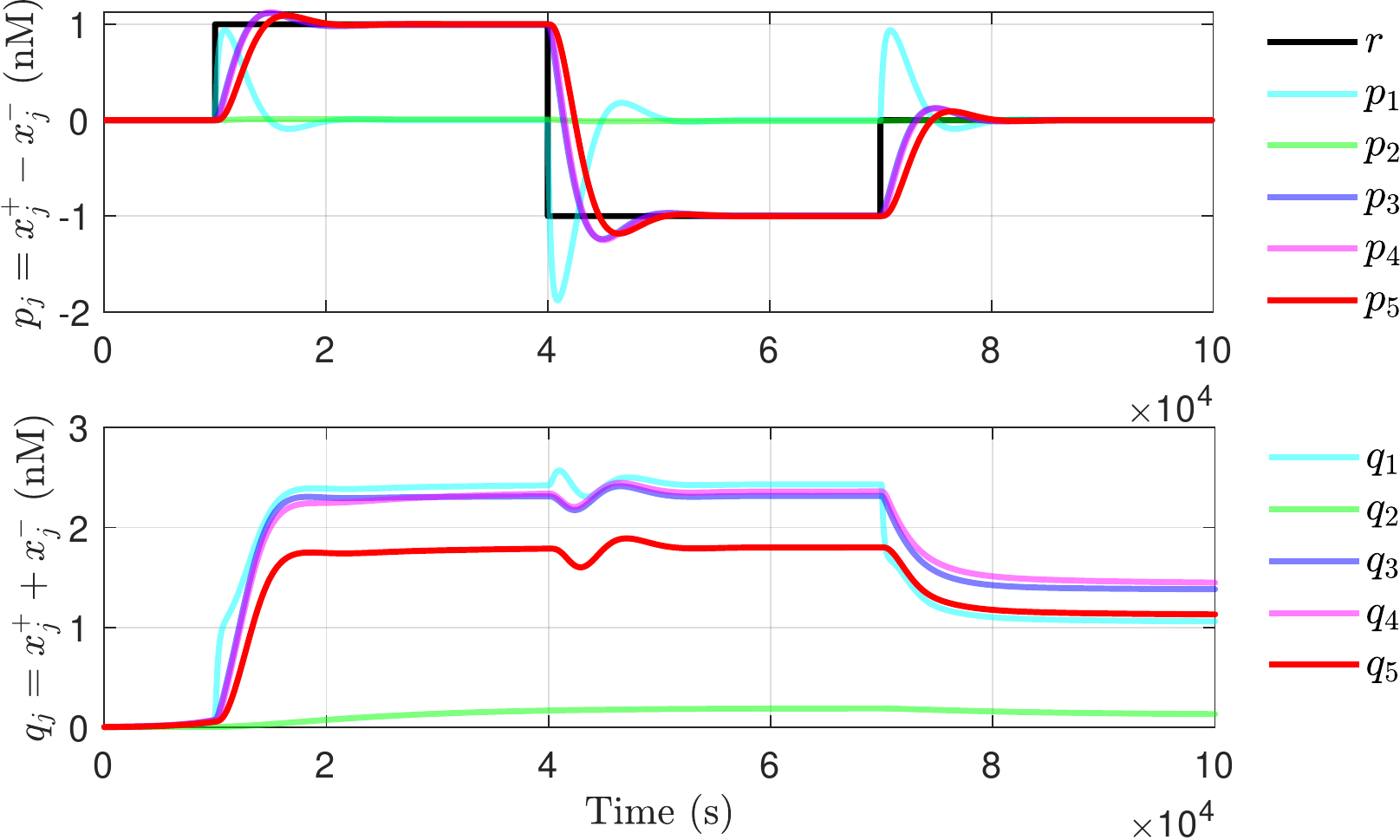}}
	\caption{Response of the CRN for the nominal parameterisation 
		($x_{j}^{\pm}(0) > 0$) to a sequence of reference steps $r$.
	}\label{Ref: fig: Nominal x}
\end{figure}

To illustrate the application of the above results, we now analyse the feedback system given in Example~\ref{exmp: ODEs CRN}. We first consider the nominal parameterisation in Table~\ref{tab: parameters}, and analyse the dynamics in the natural coordinates $x^{\pm}_j$ in~(\ref{eq: non-linear model a + A}) and the I/O linear dynamics from~\eqref{eq: I/O system state space}.
For simulation we assume that the reference signal is a sequence of steps, where only one of the concentrations $r^{+}>0$ or $r^{-}>0$ at any given time.
The response with the nominal parameterisation is shown in~Fig.~\ref{Ref: fig: Nominal x} where $\mathbf{p}$ and $\mathbf{q}$ are recovered with~\eqref{eq: Psi map}. The output $y=p_{5}$ tracks successfully the reference $r$, while $ \mathbf{q}\geq0$ reveals the underlying dynamics.
Since $\bar{\mathbf{R}}_{22}\notin\mathcal{H}$ the origin is unstable (Lemma~\ref{lem: unstable nom q=0}), and for $t>7\times 10^4\text{ s}$, when the reference returns to $r^{\pm}=0$, the state converges to a positive equilibrium  $\bar{\mathbf{x}}^{+*}=\bar{\mathbf{x}}^{-*}>0$. 

%
%

\begin{table}[h]
	\caption{Poles with maximum real part, for the I/O and linearised dynamics, for the nominal and asymmetrical parametrisations. }
	\label{tab: poles}
	\centering{}
	\begin{tabular}{p{45pt}| p{120pt} | p{36pt}}
		\hline
		Matrix $\mathbf{M}$& Poles corresponding to $\alpha\left\{\mathbf{M}\right\}$ & Stability \\
		\hline
		$ \bar{\mathbf{R}}_{11} $  &$-3.96 \times 10^{-6}$ & $\bar{\mathbf{R}}_{11}\in\mathcal{H}$\\ 
		$ \bar{\mathbf{A}}_{s}$ 	&  $-3.96\times 10^{-6}$& $ \bar{\mathbf{A}}_{s}\in\mathcal{H}$\\ \hline
		$ \mathbf{R}_{11} $  & 	 $-5.23\times 10^{-6}$& $\mathbf{R}_{11}\in\mathcal{H}$\\ 
		$\mathbf{A}_{s} $  & 
		$\mathbf{+3.16\times 10^{-5}}\pm i\mathbf{1.26\times 10^{-3}}$& $\mathbf{A}_{s}\notin\mathcal{H}$\\ 	
		\hline
		%
	\end{tabular}
\end{table}

\begin{figure}[]
	\centerline{\includegraphics[width=0.9\columnwidth]{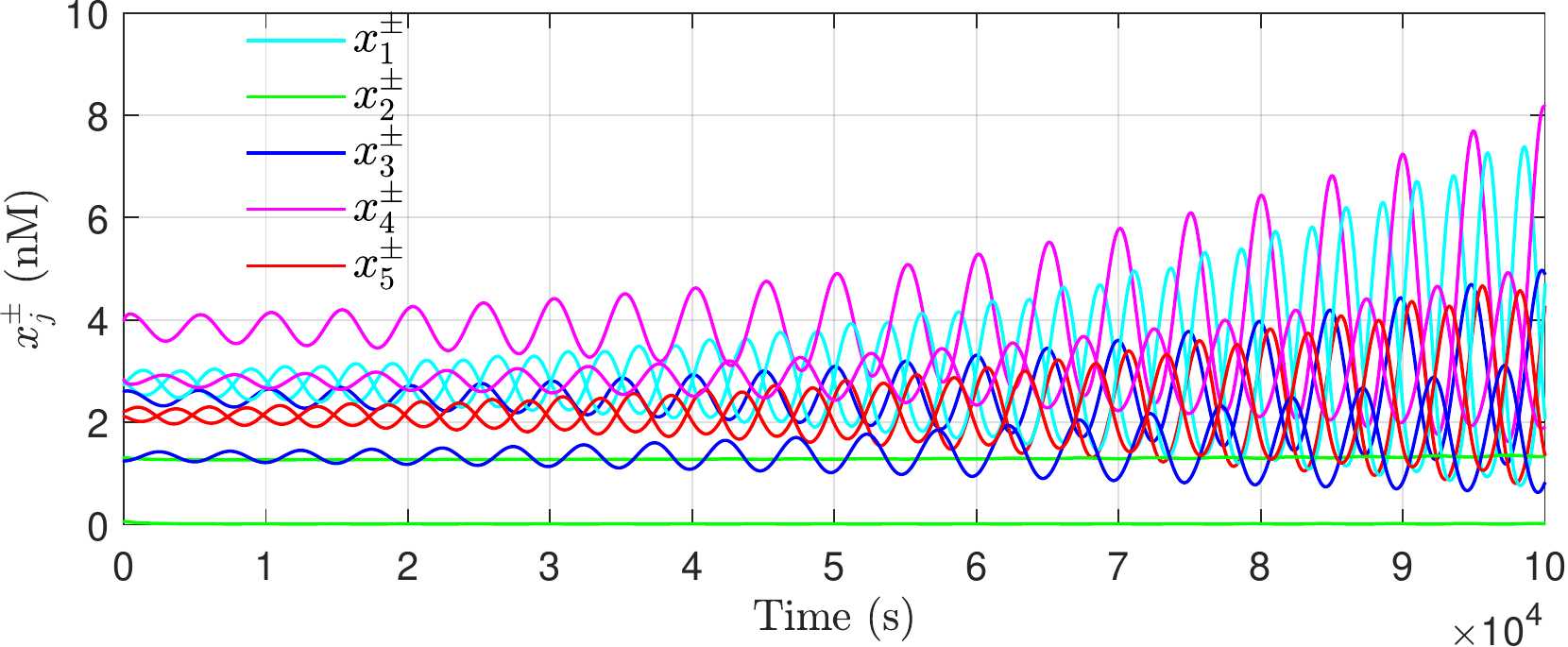}}
	\caption{Trajectories in natural coordinates $x_{i}^{\pm}$ when parameterised with the asymmetrical rates from Table~\ref{tab: parameters} ($\mathbf{r}=0$ ).
	}
	\label{fig:unstable x xi zeta}
\end{figure}

Table~\ref{tab: poles} shows that the nominal  $\bar{\mathbf{R}}_{11}$ and the linearisation around the nominal equilibrium $\bar{\mathbf{A}}_{s}$ are Hurwitz.
However, in reality, experimental variability in the reaction rates leads to asymmetric parameterisations, and the stability of I/O dynamics does not guarantee stability of the CRN.
To account for realistic levels of experimental variability, we introduced an uncertainty of $\pm 33 \%$ in the reaction rates, which includes the asymmetrical parameterisation shown in Table~\ref{tab: parameters}.
Perturbing the unforced nonlinear dynamics for this case around its equilibrium ($\mathbf{r}=0$), 
results in the unstable response of Fig.~\ref{fig:unstable x xi zeta}.
\begin{figure}[]
	\centerline{\includegraphics[width=1\columnwidth]{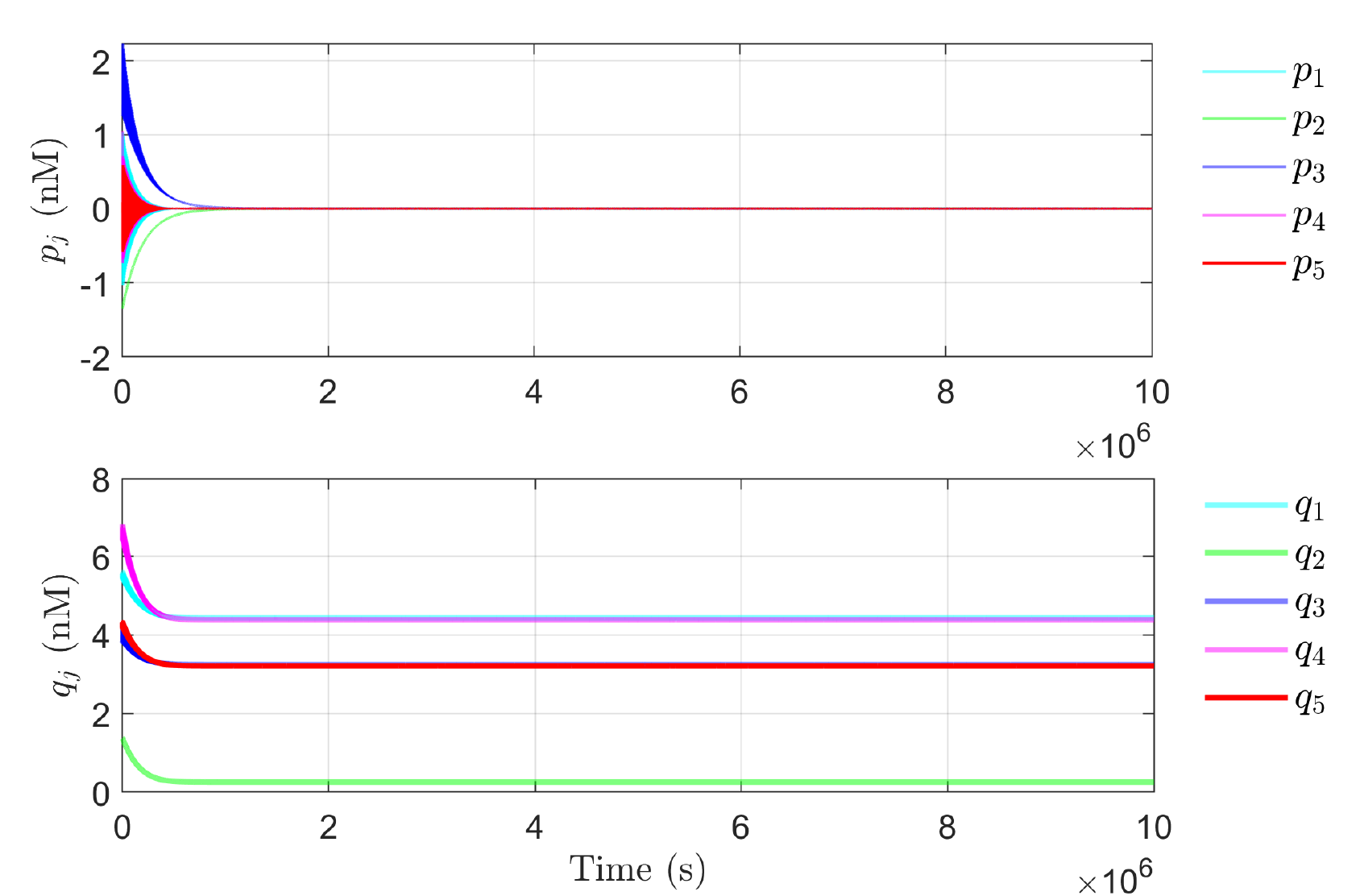}}
	\caption{Simulation of the rotated dynamics of $\dot{ \mathbf{p}}$
		and $\dot{ \mathbf{q}}$ with decoupled matrix $\mathbf{R}$ where $\mathbf{R}_{21}=\mathbf{R}_{12}=0$. 
	} \label{fig: independent xi and zeta}
\end{figure}
The poles in Table~\ref{tab: poles} show that the linearisation with the asymmetrical parameterisation  $\mathbf{A}_s$ captures the instability in a pair of conjugated poles on the right-hand plane, despite the stability of the I/O linear system $\mathbf{R}_{11}\in\mathcal{H}$. 
Indeed, integrating the rotated dynamics with a decoupled matrix $\mathbf{R}$ where we force $\mathbf{R}_{21}=\mathbf{R}_{12}=0$, we obtain the response of Fig.~\ref{fig: independent xi and zeta}, where both ${ \mathbf{p}}$ and ${ \mathbf{q}}$ have bounded trajectories. This shows that the source of the instability of the complete nonlinear system is neither $\dot{\mathbf{p}}$ nor $\dot{ \mathbf{q}}$ individually, and stability must be analysed for the complete interconnected dynamics.
%

%
%

\section{Stability of the controller implementation with DSD reactions }

It remains to verify whether the stability properties of Example~\ref{exmp: ODEs CRN} predicted from analysing the system CRNs are observed when the closed-loop system is implemented with nucleic acids.
In a DSD reaction, a strand of DNA displaces another strand from its
binding to a complementary strand
, in a random thermodynamic process, which decreases the Gibbs free energy.
The single-stranded overhangs, or toeholds, provide initial binding sites for incoming strands to initiate a toehold-mediated branch migration process that can result in strand displacement~\cite{Phillips2009,Zhang2018}. 
Tuning the affinities of the toeholds, based on the base-pair affinities and the nucleotides sequences~\cite{Zhang2018}, allows the mapping of the desired reaction rates for the CRN into the DSD implementation.

Following~\cite{Soloveichik2010}, the chemical reactions result in bimolecular DSD reactions, which produce \emph{waste} in the form of inactivated double strands of DNA which cannot participate in any reaction.
Auxiliary \emph{fuel} species are consumed irreversibly as fuel, and the reactions stop if these are not replenished (details in \cite{Soloveichik2010}). 
The fuel species are initialised at a high concentration $C_{max}=10^{4}\,\text{nM}$, to prevent their consumption from impacting the dynamics significantly.

\begin{figure}[t]
	\centering
	\includegraphics[width=1\columnwidth]{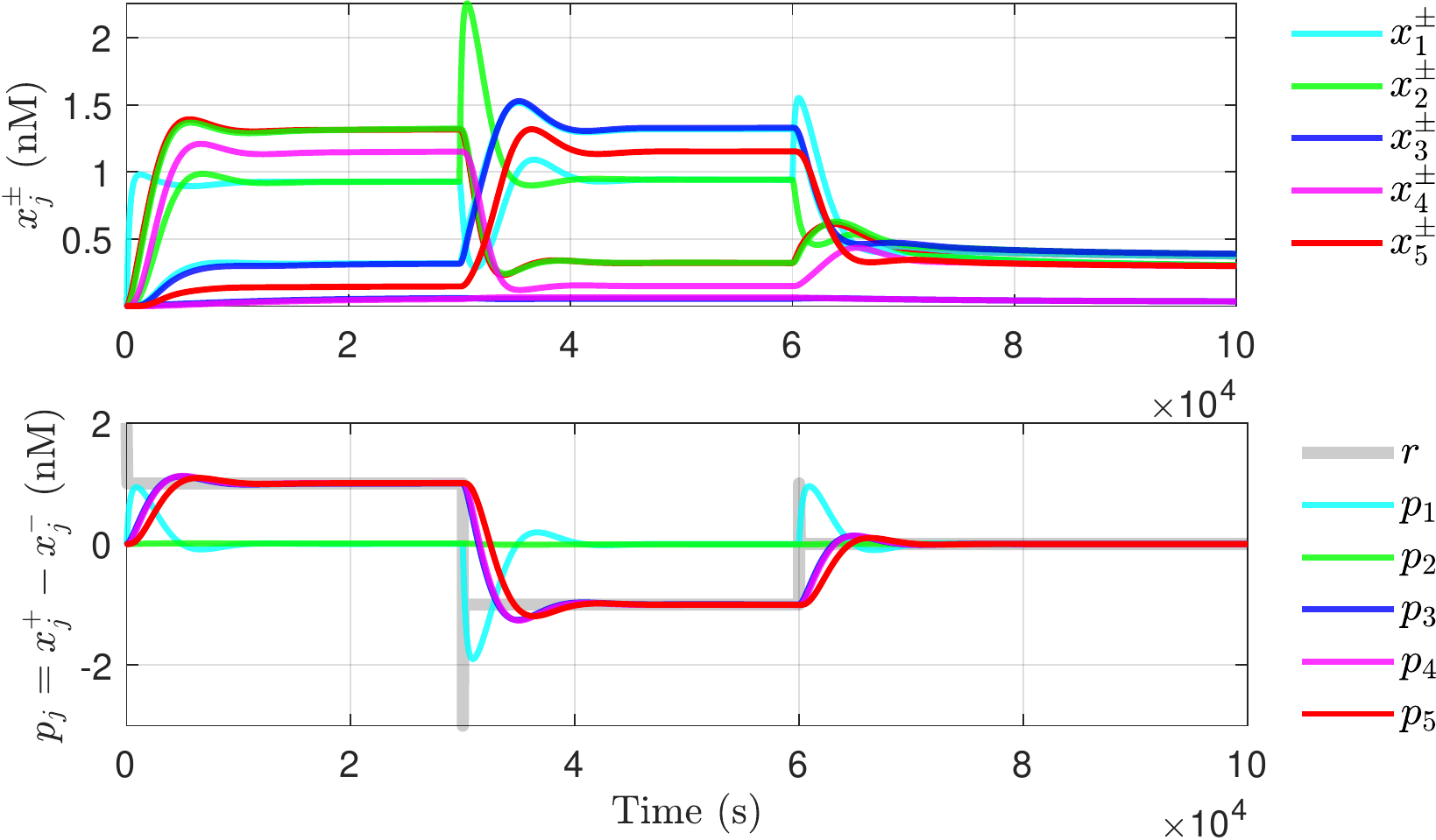}
	\caption{Simulation of the DSD reactions in VisualDSD for the symmetrical nominal system, with $\mathbf{x}\left(0\right)=0 $ nM and a sequence of steps on $\mathbf{r}$. 
	}\label{fig: DSD nominal} 
\end{figure}
\begin{figure}[t]
	\centerline{\includegraphics[width=0.95\columnwidth]{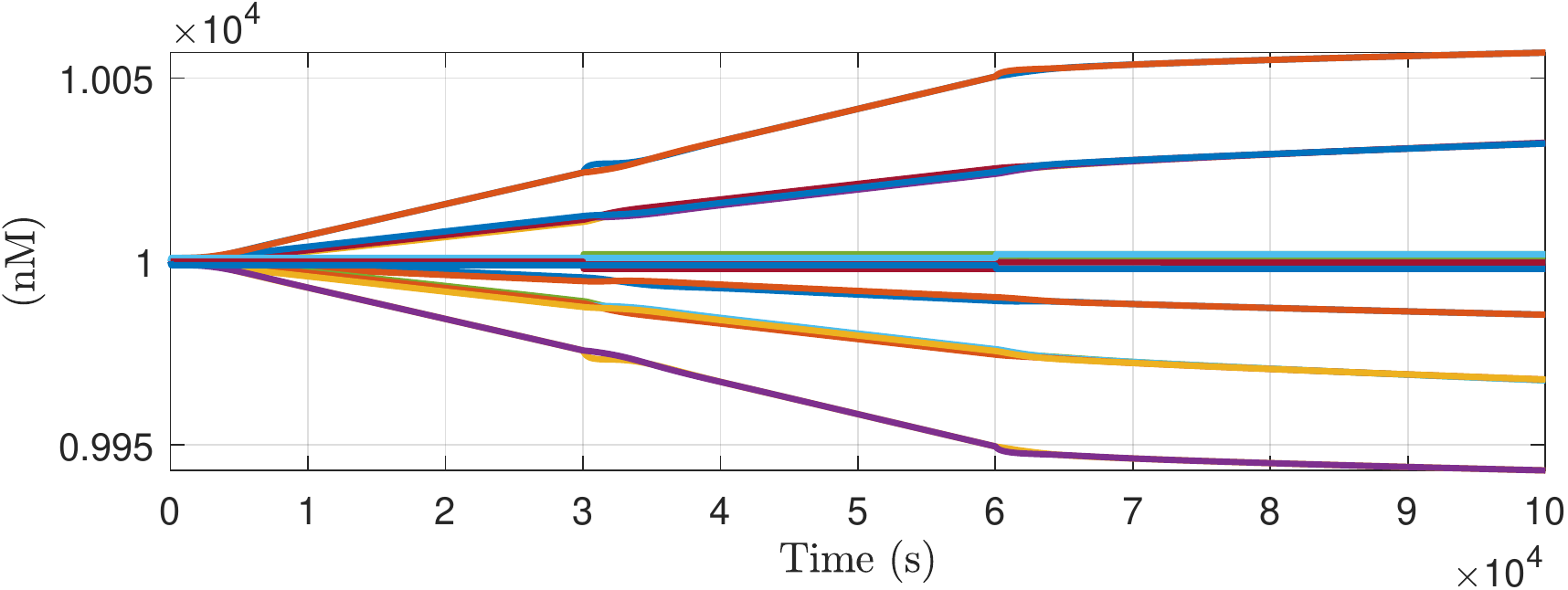}}
	\caption{Time history of concentrations of auxiliary species involved in the annihilation reactions. 
		\label{fig: DSD nominal aux} }
\end{figure}
\begin{figure}[t]
	\centering
	\includegraphics[width=0.95\columnwidth]{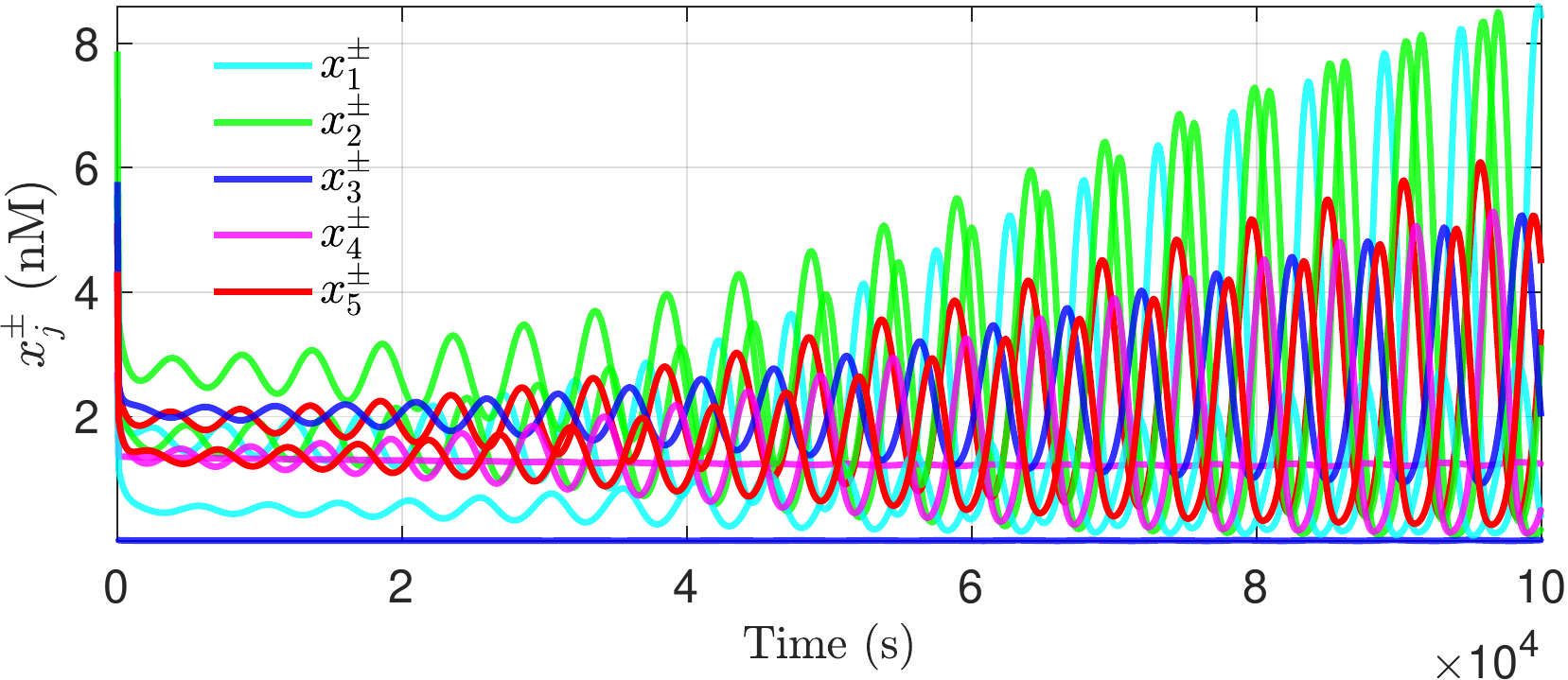}
	\caption{Simulation in VisualDSD, for the asymmetrical destabilising parameterisation, with $\mathbf{r}=0$.
		\label{fig: DSD ucase}} 
\end{figure}
The DNA strand displacement reactions are simulated using VisualDSD, a rapid-prototyping tool that allows precise analysis of computational devices implemented using DNA strand displacement reactions~\cite{lakin2011visual}.
The translation of the CRN system follows the construction proposed in~\cite{Soloveichik2010}, with a two-domain programmming structure~\cite{cardelli2013two}. The model was parameterised with the unstable parameterisation of~Table~\ref{tab: parameters}, applying the correspondence between the reaction rates in the CRN and the DSD implementation: $q_{max}=2\eta$, 
$q_{ki}^\pm=2k_i^\pm/C_{max}$,  $i\in \{0,1,2\}$, and $q_i^\pm=2\gamma_i^\pm/C_{max}$, $i\in \{1,\dots,8\}$. 

The behaviour of the nominal symmetrical parameterisation is first verified in Fig.~\ref{fig: DSD nominal}, where $p_5$ tracks the step inputs of $r$. After $6\times 10^{6}$ s, the system converges to the positive equilibrium, and Fig.~\ref{fig: DSD nominal aux} shows the concentrations of the auxiliary species involved in the annihilation reactions remain around $C_{max}$ but are still depleted when $r^\pm=p_j=0$.
Fig.~\ref{fig: DSD ucase} shows that the parameterisation which destabilises the CRN also destabilises the DSD implementation, emphasising the practical relevance of the stability results.

%
%

\section{Conclusions }

Several recent works have applied the dual-rail representation of CRN's to obtain linear I/O models of synthetic feedback control systems, but have not explicitly considered the potential impact of the underlying nonlinear annihilation reactions in their analysis.
This new class of IPR derived from CRNs relies on internally nonlinear positive dynamics.
%
We decomposed the dynamics of the CRN's involved in a typical linear controller design, and highlighted the effects of the non-observable and nonlinear dynamics - in particular, we showed that the stability of these I/O models does not imply the stability of the underlying chemical network.
Under inevitable experimental variability, stability can be affected by the looped interconnection between the nonlinear dynamics arising from biochemical implementation and the linear I/O dynamics resulting from the controller designs. 
We presented an example of this phenomenon, where the I/O linear system does not capture the instability of the full nonlinear system, and verified this result via simulation of the DSD network that would be implemented experimentally. Our results confirm that the stability of nucleic acid-based controllers must be analysed using the linearisation of the complete nonlinear system, and provide a rigorous theoretical approach for conducting such an analysis.

\section*{Acknowledgements }

DGB acknowledges funding from the University of Warwick, the EPSRC/BBSRC Centre for Doctoral Training in Synthetic Biology via grant EP/L016494/1 and the BBSRC/EPSRC Warwick Integrative Synthetic Biology Centre via grant BB/M017982/1.

\appendix


\bibliographystyle{unsrt}        
\bibliography{library}           

\balance

\ifarxiv
\else
\end{document}
\fi

\cleardoublepage

\section{The representation of cascaded dynamics has a single equilibrium}

For cascaded systems $\bar{\mathbf{R}}_{22}\in\mathcal{L}$, and we can apply Lemma~\ref{lem: equilibrium L nominal} to conclude that the unforced equilibrium of the cascaded dynamics  $\dot{\mathbf{q}}=\bar{\mathbf{R}}_{22}\mathbf{q}-\frac{\eta}{2}\mathbf{q}\circ\mathbf{q}$ is unique at $\mathbf{q}=0$.
\begin{lemma}\label{lem: equilibrium L nominal}
	If $\mathbf{M}\in\mathcal{H}$ is also a lower triangular matrix $\mathbf{M}\in\mathcal{L}$, then $\dot{ \mathbf{q}}=\mathbf{Mq}-k\mathbf{q}\circ \mathbf{q}$ has a single equilibrium $\mathbf{q}=0$.
	%
\end{lemma}
\begin{pf}
	Given $\mathcal{M}\in\mathcal{L}$, the solution for $q_j$ depends only on $q_i,\,i\leq j$, with
	\begin{eqnarray}
	q_{j}^{2}-m_{jj}q_{j}-\sum_{i=1}^{j-1}m_{ji}q_{i}=0
	\end{eqnarray}
	For $j=1$, the solutions are $q_{1}\in\{0,m_{11}\}$.
	$\mathcal{M}\in\mathcal{H}, \mathcal{L}\Rightarrow m_{11}\leq 0$ and the only non-negative solution is $q_{1}=0$. 	Solving sequentially for the remaining coordinates $j=2,3,\ldots,N$, knowing that $q_{i<j}=0$, we
	have that $\sum_{i=1}^{j-1}m_{ji}q_{i}=0\Rightarrow q_{j}\in\{0,m_{jj}\}$. Since $m_{jj}\leq 0$, the non-negative solution is always $q_j=0$.\qed
\end{pf}

\section{Stability analysis and positive equilibrium conditions for Example~\ref{exmp: 2 state example}}

Following on from Remark~\ref{prop: negative vs positive feedback}, we see how $c_2$ impacts differently the spectral radius of $\bar{\mathbf{R}}_{11}$ and $\bar{\mathbf{R}}_{22}$. The characteristic polynomial for the I/O dynamics
\begin{eqnarray}
&&\lambda\left\{ \mathbf{R}_{11}\right\} :\lambda^{2}+\lambda\left(d_{1}+d_{2}\right)+d_{1}d_{2}+c_{1}c_{2} =0 \label{eq: ch.poly.R11}\\
&&\Rightarrow \lambda=-\frac{\left(d_{1}+d_{2}\right)}{2}\pm\frac{1}{2}\sqrt{\left(d_{1}+d_{2}\right)^{2}-4d_{1}d_{2}-4c_{1}c_{2}}\nonumber
\end{eqnarray}
shows that the I/O system is stable for any $c_{2}>0$.
%
On the other hand
\begin{eqnarray}
&&\lambda\left\{ \bar{\mathbf{R}}_{22}\right\} :\lambda^{2}+\lambda\left(d_{1}+d_{2}\right)+d_{1}d_{2}-c_{1}c_{2} =0 \Rightarrow\label{eq: ch.poly.R22}\\
&&\lambda   =-\frac{\left(d_{1}+d_{2}\right)}{2}\pm\frac{1}{2}\sqrt{\left(d_{1}+d_{2}\right)^{2}+4\left(d_{1}d_{2}-c_{1}c_{2}\right)}\nonumber
\end{eqnarray} 
and a gain $ c_{2}>{d_2 d_1}{c_1^{-1}}$ which stabilises the linear I/O dynamics leads to $\bar{\mathbf{R}}_{22}\notin\mathcal{H}$. 
Furthermore, the domain for which $\bar{\mathbf{R}}_{22}\notin\mathcal{H}$ and $\mathbf{q}^*>0$ exists is the same:  $c_{2}>{d_2 d_1}{c_1^{-1}}$. %

\section{Representation of the stability bounds}

\begin{figure}[t]
	\begin{centering}
		\includegraphics[width=.9\columnwidth]{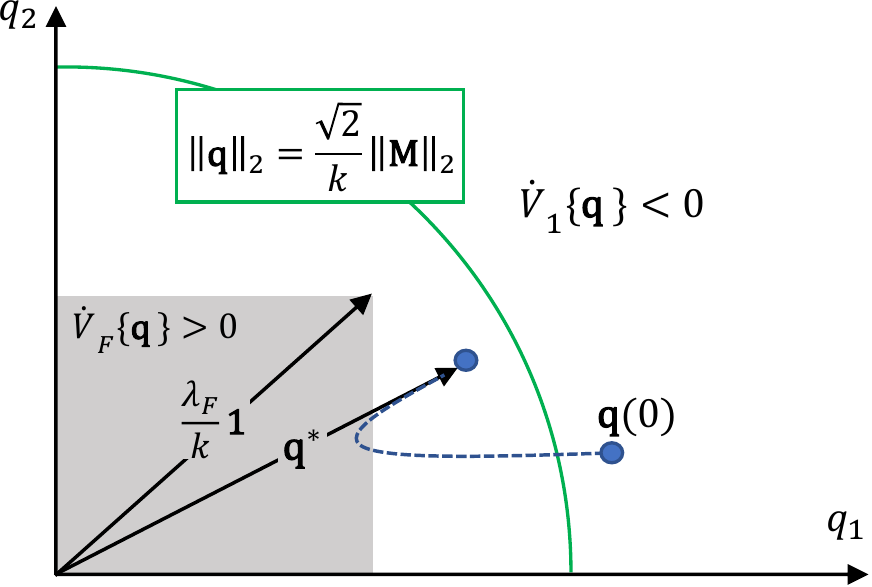}
		\par\end{centering}
	\caption{Illustration of the exclusion area from Lemma~\ref{lem: unstable nom q=0} (gray), and the upper bound from Lemma~\ref{lem: bounded nominal q} (green), for the trajectories of the nominal nonlinear dynamics of a 2 dimensional system. Considering $\mathbf{M}\in\mathcal{I},\mathcal{M}$ and $\mathbf{M}\notin\mathcal{H}$,  the trajectories close to the origin will diverge. The dashed curve is an illustration of convergence to the equilibrium ${\mathbf{q}}^{*}$.
		\label{fig: q0 Vdot}}
\end{figure}

Fig.~\ref{fig: q0 Vdot} illustrates the stability results from Lemmas~\ref{lem: unstable nom q=0} and~\ref{lem: bounded nominal q}.

\section{Stability of the CRN representation for a cascaded system, under parameter variability}

We can easily state a stability condition for the representation of a cascaded system, even if experimental variability results in an asymmetrical parameterisation.
\begin{proposition}\label{prop: stab cascaded asym}
Take the representation of a stable cascaded system $\dot{\mathbf{p}}=\bar{\mathbf{R}}_{11}\mathbf{p}, \bar{\mathbf{R}}_{11}\in\mathcal{L},\mathcal{H}$. For an asymmetrical parameterisation (without Assumption~\ref{assum: rates are equal}), if $\mathbf{a}_{1}^{\pm}<0$, the unforced dynamics $\dot{\mathbf{x}}=\mathbf{Ax}-\eta\mathbf{x}\circ\left(\mathbf{Px}\right)$ are GAS for $\mathbf{x}=0$.
\end{proposition}
\begin{pf} 
	Given a cascaded I/O dynamics, then we can permutate the state $\mathbf{p}$ so that $\bar{\mathbf{R}}_{11}\in\mathcal{L}$, resulting also $\bar{\mathbf{A}}_{1},\bar{\mathbf{A}}_{2}\in\mathcal{L}$. 
	In the presence of variability, $\mathbf{A_{1}^{\pm}}$ have the same structure as $\bar{\mathbf{A}}_{1}$ but with different parameterisations, resulting $\mathbf{A_{1}^{\pm}}\in\mathcal{L}$. In the same way, $\mathbf{A}_{2}^{\pm}\in\mathcal{L}$.
Now take the permutation matrix $\mathbf{Q}$ 
\begin{eqnarray}
&&\mathbf{Q}=\left[\begin{array}{cccc|cccc}
1 & 0 & \ldots & 0 & 0 & 0 & \dots & 0\\
0 & 0 & \ldots & 0 & 1 & 0 & \ldots & 0\\\hline
0 & 1 & \dots & 0 & 0 & 0 & \ldots & 0\\
0 & 0 & \ldots & 0 & 0 & 1 & \dots & 0\\\hline
\vdots & \vdots & \ddots & \vdots & \vdots & \vdots & \ddots & \vdots\\\hline
0 & 0 & \ldots & 1 & 0 & 0 & \ldots & 0\\
0 & 0 & \ldots & 0 & 0 & 0 & \ldots & 1
\end{array}\right]
\end{eqnarray}
such that
\begin{eqnarray}
&&\mathbf{z}=\mathbf{Q}\mathbf{x}=\left[\begin{array}{cc|cc|c|cc}
x_{1}^{+} & x_{1}^{-} & x_{2}^{+} & x_{2}^{-} & \dots & x_{N}^{+} & x_{N}^{-}\end{array}\right]^{T}
\end{eqnarray} 
The dynamics of the permuted state result
\begin{eqnarray}
&&\dot{\mathbf{z}}=\mathbf{L}\mathbf{z}-\eta\mathbf{z}\circ\mathbf{g}\left\{ \mathbf{z}\right\}
\end{eqnarray}
where $\mathbf{z} >0\Rightarrow \mathbf{g}\left\{ \mathbf{z}\right\}<0$ and
\begin{eqnarray}
&& \mathbf{L}=\left[\begin{array}{cc}
\mathbf{L}_{11} & \mathbf{0}\\
\mathbf{L}_{21} & \mathbf{L}_{22}
\end{array}\right]
\end{eqnarray}
The structures of $\mathbf{L}_{jj}$ are determined by the structure of $\mathbf{A}_1^\pm$, and $\mathbf{L}_{21}\geq0$ contains the cross terms which result in subtractions in the I/O dynamics (elements in $\mathbf{A}_2^\pm$).
Since $\mathbf{A_{1}^{\pm}},\mathbf{A_{2}^{\pm}}\in\mathcal{L}$, $\mathbf{L}$ results
triangular,  Moreover, $\mathbf{L}=\mathbf{L}^{\boxbslash}+\mathbf{D}\left\{ \mathbf{l}\right\} $ where  $\mathbf{l}=\mathbf{Q}\left[\begin{array}{cc}
\mathbf{a}_{1}^{+T} & \mathbf{a}_{1}^{-T}\end{array}\right]$. 

It results directly that $\rho\left\{ \mathbf{A}\right\} =\rho\left\{ \mathbf{L}\right\}=\rho\left\{ \mathbf{A}_{1}^{+}\right\} \cup\rho\left\{ \mathbf{A}_{1}^{-}\right\} $, and  $\mathbf{a}_{1}^\pm<0\Leftrightarrow\mathbf{L}\in\mathcal{H}\Leftrightarrow\mathbf{A}\in\mathcal{H}$. If the represented cascaded linear I/O dynamics are stable, then $\mathbf{A}\in\mathcal{H}$. Moreover, even with uncertainty, as long as the degradation rates remain positive $\mathbf{A}\in \mathcal{H}$.
Since $\mathbf{A}\in\mathcal{H}$, we can invoke Lemma~\ref{lm: stable if M hurwitz} to establish $\dot{\mathbf{x}}=\mathbf{A}\mathbf{x}-\eta\left(\mathbf{Px}\right)\circ\mathbf{x}$ is GAS around $\mathbf{x}=0$. \qed
\end{pf}

In the presence of variability we can have mismatching rates, but as long as all species degrade with some non-zero rate, the unforced dynamics of the cascaded system will have a single stable nonnegative equilibrium. Without input, the CRNs will converge to rest at $\mathbf{x}=0$.
Table~\ref{tab: summary table} summarises the derived properties, depending on the structure of the DSD network (cascaded \emph{versus} with feedback).

\begin{table}[t]\label{tab: summary table}
	\begin{tabular}{|p{50pt}|p{75pt}|p{75pt}|}
		\hline 
		Parameters & Cascaded & With feedback\tabularnewline
		\hline 
		\hline 
		Nominal \par $\bar{\mathbf{R}}_{11}\in\mathcal{H}$  & $\mathbf{x}^{*}=0$ 
		\par Unforced dynamics are GAS & Possible $\mathbf{x}^{*}>0$ \par 		
		Unforced dynamics are bounded \tabularnewline
		\hline 
		Asymmetrical \par  $\mathbf{R}_{11}\in\mathcal{H}$ &  $\mathbf{x}^{*}=0$  \par Unforced dynamics are GAS if additionally $\mathbf{a}_1^\pm<0$  & Possible $\mathbf{x}^{*}>0$ \par		
		CRN may be unstable \tabularnewline
		\hline 
	\end{tabular}%
	\caption{Assuming the I/O system is stable, we can state properties about CRN stability and the unforced equilibria $\mathbf{x}^*$.}
\end{table}
\end{document}